\documentclass[aps, prb,reprint,10pt,superscriptaddress,amssymb,amsfonts]{revtex4-1}
\usepackage{amsmath,amssymb,amsthm,amscd,latexsym,epsfig,bm}

\usepackage[colorlinks,bookmarks=false,citecolor=blue,linkcolor=red,urlcolor=blue]{hyperref}
\usepackage{verbatim}
\usepackage[normalem]{ulem}
\usepackage[usenames, dvipsnames]{color}

\def\beq{\begin{equation}}
\def\eeq{\end{equation}}
\def\be{\begin{equation}}
\def\ee{\end{equation}}

\newcommand{\zz}{\mathbb{Z}_2}
\newcommand{\z}{\mathbb{Z}}
\newcommand{\uzz}{{\rm U}(1)-\mathbb{Z}_2}

\def\bS{\boldsymbol{S}}

\def\f2{{\mathbb F}_2}

\theoremstyle{plain}
\theoremstyle{plain}

\providecommand{\theoremname}{Theorem}
\providecommand{\theoremtextname}{Theorem}

\theoremstyle{plain}
\providecommand{\propositionname}{Proposition}

\begin{document}

\title{Surface field theories of point group symmetry protected topological phases}
\author{Sheng-Jie Huang}
\affiliation{Department of Physics, University of Colorado, Boulder, Colorado 80309, USA}
\affiliation{Center for Theory of Quantum Matter, University of Colorado, Boulder, Colorado 80309, USA}
\author{Michael Hermele}
\affiliation{Department of Physics, University of Colorado, Boulder, Colorado 80309, USA}
\affiliation{Center for Theory of Quantum Matter, University of Colorado, Boulder, Colorado 80309, USA}
\date{\today}

\begin{abstract}

We identify field theories that describe the surfaces of three-dimensional bosonic point group symmetry protected topological (pgSPT) phases. The anomalous nature of the surface field theories is revealed via a dimensional reduction argument. Specifically, we study three different surface field theories. The first field theory is quantum electrodynamics in three space-time dimensions (QED3) with four flavors of fermions.  We show this theory can describe the surfaces of a majority of bosonic pgSPT phases protected by a single mirror reflection, or by $C_{nv}$ point group symmetry for $n=2,3,4,6$. The second field theory is a variant of QED3 with charge-1 and charge-3 Dirac fermions. This field theory can describe the surface of a reflection symmetric pgSPT phase built by placing an $E_{8}$ state on the mirror plane. The third field theory is an ${\rm O}(4)$ non-linear sigma model with a topological theta-term at $\theta=\pi$, or, equivalently, a non-compact ${\rm CP}^1$ model.  Using a coupled wire construction, we show this is a surface theory for bosonic pgSPT phases with ${\rm U}(1) \times \zz^{P}$ symmetry.  For the latter two field theories, we discuss the connection to gapped surfaces with topological order.  Moreover, we conjecture that the latter two field theories can describe surfaces of more general bosonic pgSPT phases with $C_{nv}$ point group symmetry.

\end{abstract}

\maketitle

\section{Introduction}

Symmetry protected topological (SPT) phases\cite{kitaev2009, ryu2010, pollmann2010, fidkowski2011, turner2011, chen2011_1dspt, chen2011_1dcomplete, cirac2011, chen2013cohomology, levin2012} are a class of gapped phases of matter, which behave almost trivially in the bulk, but support interesting boundary states. SPT phases, by definition, have a bulk energy gap, a unique ground state for periodic boundary conditions, and can be adiabatically connected to a trivial product state if the symmetry is broken explicitly. Topological band insulators\cite{hasan2010review, qi2011review, hasan2011review} serve as a key example of an SPT phase, where the symmetries needed to protect the non-trivial boundary states are $\z_2^T$ time reversal and ${\rm U}(1)$ charge conservation.  These are both internal symmetries, and more generally there has been remarkable progress in the characterization and classification of SPT phases protected by internal symmetry.\cite{senthil2015review, chiu2016review}

Besides internal symmetries, crystalline symmetries such as point group or space group symmetry can also protect SPT phases.  Such crystalline SPT (cSPT) phases, including topological crystalline insulators (TCIs),\cite{fu2011tci, fu2015review} are rather well-understood in non-interacting fermion systems.\cite{chiu2016review}  When interactions are strong, less is known about cSPT phases than their internal symmetry cousins. However, a number of works have studied examples of interacting cSPT phases,\cite{chen2011_1dspt, chen2011_1dcomplete, hsieh2014, you2014, hsieh2014cpt, isobe15, cho2015, yoshida2015, kapustin2015, sid2015, qi2015anomalous, yoshida2015tci, morimoto2015, fuji2015, lapa2016, hsieh2016, lake2016, schnyder2017} and significant progress on general classification has been made recently.\cite{song2017,jiang17anyon,thorngren16gauging,huang17building}  In particular, it has been shown that point group SPT (pgSPT) phases can be built by placing lower-dimensional topological states at points, lines and planes of high symmetry.\cite{song2017}  Crucial to this result is a dimensional reduction procedure, where a pgSPT wave function is locally trivialized away from a lower-dimensional subspace containing points fixed under symmetry operations.  While this approach does not apply directly in the presence of space group symmetry, many space group SPT phases, and perhaps all cSPT phases, can nonetheless be constructed and classified in a similar fashion.\cite{huang17building}  Other approaches to the classification of cSPT phases are based on an extension of the notion of gauging symmetry to crystalline symmetries,\cite{thorngren16gauging} and on tensor-network wave functions.\cite{jiang17anyon}  These approaches give identical classifications, at least for bosonic states in $d=2,3$ protected only by crystalline symmetry and built from lower-dimensional SPT states in the dimensional reduction approach.\cite{huang17building}

While Ref.~\onlinecite{song2017} used dimensional reduction in the bulk to classify and characterize pgSPT phases, the procedure can also be used on the surface.  Indeed, the approach of Ref.~\onlinecite{song2017} was inspired in part by the work of Isobe and Fu,\cite{isobe15}  who used a surface version of dimensional reduction to show that the $\z$ classification of non-interacting electron TCIs protected by mirror reflection reduces to $\z_8$ in the presence of interactions. 
 In this paper, we show that dimensional reduction is a simple means of identifying and characterizing surface field theories of pgSPT phases.   In particular, we focus on pgSPT phases in spatial dimension $d=3$ and their two-dimensional surfaces. We study three different field theories, showing that two of these are ``parent theories,'' in the sense that they can describe multiple different pgSPT surfaces, 
 depending on the microscopic symmetry and how it acts on the continuum fields.  The other theory is conjectured but not shown to describe a number of different pgSPT surfaces.
 
In general, SPT surfaces are anomalous, in the sense that they possess properties that are intrinsically due to the existence of a bulk.  A familiar example is the single Dirac cone observed on the surface of $d=3$ topological insulators,\cite{hasan2010review, qi2011review, hasan2011review, fu20073dti} which cannot occur in a purely $d=2$ system with time reversal symmetry. The surface dimensional reduction we employ here can be viewed as a tool to reveal anomalies in a field theory, and to match it to a corresponding pgSPT bulk.

We study three different surface field theories for bosonic pgSPT phases.  Two of these are also related to fermion pgSPT surface theories, and are variants of quantum electrodynamics in three space-time dimensions (QED3), where Dirac fermions are coupled to a compact ${\rm U}(1)$ gauge field.  The first variant is QED3 with $N_f = 4$ fermion fields, which can be obtained from the surface of an electron TCI with four Dirac fermions by gauging the ${\rm U}(1)$ conserved charge.  We show this QED3 theory can describe surfaces of all bosonic pgSPT phases built either by placing $d=2$ Ising SPT states\cite{chen2013cohomology, levin2012} on mirror planes, or by placing Haldane chains\cite{haldane1983, haldane1983nlsm, affleck1987} on $C_{nv}$ axes.

The second variant of QED3 has a single charge-1 and a single charge-3 Dirac fermion, and can be obtained, by gauging ${\rm U}(1)$ symmetry, from the surface of the strongly interacting $E_8$ paramagnet TCI discussed in Ref.~\onlinecite{song2017}.  We arrive at this theory by combining dimensional reduction with the ``cluston'' constructions of various interacting topological phases introduced in Ref.~\onlinecite{chong2015}, where clustons are fermions formed by binding together an odd number of electrons (see also Ref.~\onlinecite{seiberg2016} for a related discussion). This QED3 theory describes the surface of the mirror-symmetric bosonic pgSPT phase obtained by placing an $E_8$ state\cite{kitaev11KITP} on the mirror plane.  We also conjecture that the same theory describes surfaces of more general pgSPT phases obtained by placing $E_8$ states on mirror planes, but we do not establish this definitively.

The third field theory we study for bosonic pgSPT surfaces is the ${\rm O}(4)$ non-linear sigma model with a topological theta-term at $\theta = \pi$, which can be mapped to the ${\rm CP}^1$ model of a two-component bosonic field coupled to a ${\rm U}(1)$ gauge field.\cite{senthil2006} We focus on bosonic pgSPT phases protected by ${\rm U}(1) \times \zz^{P}$ symmetry, although we conjecture that this theory can describe surfaces of more general bosonic cSPT phases with symmetry ${\rm U}(1) \times G_c$, where $G_c$ is a point group or space group.
 
The field theories considered in this paper have been discussed as surface theories of SPT phases involving time-reversal symmetry.\cite{ashvin2013, chong2014, witten2016, hsieh2016, chong2015, seiberg2016, you2014, you2015, bi2015classification} Indeed, these theories are Lorentz-invariant, and if we work in Euclidean space-time, there is no difference between reflection and time-reversal symmetry. A reflection symmetry in a Euclidean field theory can be analytically continued to either a reflection or a time-reversal symmetry in Lorentz signature (see Ref.~\onlinecite{witten2016} for a discussion of this well-known fact in the context of topological phases).  From this point of view, the dimensional reduction method can serve as an alternative and particularly simple means to obtain surface field theories for SPT phases involving time-reversal symmetry.

We now give an outline of the paper. We first review the classification of $d=3$ pgSPT phases in Sec.~\ref{sec:dimr}, following Ref.~\onlinecite{song2017}.  We focus on point groups that can be preserved at surfaces; among such point groups, there are non-trivial pgSPT phases with $C_s$ and $C_{nv}$ symmetry ($n=2, 3, 4, 6$). These states can be obtained by placing Ising SPT or $E_{8}$ states on mirror planes, and by placing Haldane chains on $C_{nv}$ axes. Sec.~\ref{sec:dirac} reviews the Dirac surface theory of TCIs with ${\rm U}(1) \times \zz^{P}$ symmetry. Surface dimensional reduction is discussed and used as a tool to reveal anomalies of the Dirac theory.

Sec.~\ref{sec:4dirac} considers QED3 with $N_{f}=4$ Dirac fermions as a surface field theory of bosonic pgSPT phases. Sec.~\ref{sec:4dirac_tci} focuses on bosonic pgSPT phase with a single reflection symmetry, built by placing an Ising SPT state on the mirror plane. The QED3 surface theory is obtained from the Dirac surface theory of an $n=4$ electronic TCI with ${\rm U}(1) \times \zz^{P}$ symmetry, by gauging the ${\rm U}(1)$ symmetry.  The connection to the $N_{f}=4$ QED3 theory obtained by Qi and Fu\cite{qi2015anomalous} is also discussed in this section and in Appendix~\ref{app:particle-hole}. We further discuss how to modify the symmetry action in the $N_{f}=4$ QED3 theory such that it describes a surface of a trivial bosonic pgSPT state. Sec.~\ref{sec:4dirac_c2v} considers bosonic pgSPT phases with $C_{2v}$ symmetry, built by placing Ising SPT states on mirror planes. Sec.~\ref{sec:cnv} extends the discussion to $C_{nv}$ symmetry. Sec.~\ref{sec:1d_haldane} considers the surfaces of the remaining pgSPT phases built by placing a Haldane chain on the $C_{nv}$ axis. Taken together, these results show that $N_{f}=4$ QED3 theory can describe surfaces of all bosonic pgSPT phases built either from lower-dimensional SPT states placed on high-symmetry planes and axes.

Sec.~\ref{sec:e8} discusses the surface field theories of bosonic and fermionic SPT states built by placing an $E_{8}$ state on the mirror plane. We first consider an $E_{8}$ based electronic TCI, dubbed the $E_{8}$ paramagnet TCI.\cite{song2017} The surface field theory is obtained by the cluston construction,\cite{chong2015} and is a Dirac theory with charge-1 and charge-3 Dirac fermions. The surface theory of the $E_{8}$ based bosonic pgSPT phase is obtained by gauging the ${\rm U}(1)$ symmetry. This field theory is a variant of QED3 with both charge-1 and charge-3 Dirac fermions. We then discuss how to obtain a gapped surface with three-fermion topological order starting from the QED3 field theory.

Sec.~\ref{sec:nlsm} considers a parent surface field theory of some bosonic pgSPT phases with ${\rm U}(1) \times \zz^{P}$ symmetry, which we classify in Appendix~\ref{app:pgspt_u1m}. The surface field theory is an ${\rm O}(4)$ non-linear sigma model with a topological theta-term at $\theta=\pi$ and with ${\rm O}(2) \times {\rm O}(2)$ anisotropy, which is equivalent to a non-compact ${\rm CP}^{1}$ model.\cite{senthil2006} This surface field theory is obtained by a coupled wire construction where each wire is a ${\rm SU}(2)_{1}$ Wess-Zumino-Witten model. The connection to the surface topological orders is also discussed. Finally, we conclude in Sec.~\ref{sec:discussion} with a discussion of our results and possible directions for future work.

\section{Review}
\label{sec:review}

\subsection{Point group SPT phases built from lower-dimensional states}
\label{sec:dimr}

We begin by reviewing results of Refs.~\onlinecite{song2017,huang17building} on classification of pgSPT phases in terms of lower-dimensional topological states. Ref.~\onlinecite{song2017} studied pgSPT phases in $d$ spatial dimensions, and argued that a ground state pgSPT wave function can be adiabatically connected to a product of a trivial state with a non-trivial wave function on a lower-dimensional subset of $d$-dimensional space.  A particularly simple example is mirror reflection symmetry in three-dimensions, which is the point group $C_s$.  We also refer to this point group as $\z_2^P$.  Here, the lower-dimensional space is simply the $d=2$ mirror plane.  On the mirror plane, reflection acts as a $\z_2$ effective internal symmetry, which allows one to classify $d=3$ mirror-symmetric pgSPT phases by placing $d=2$ topological states with $\z_2$ symmetry on the mirror plane.  In bosonic systems, there are two root states that generate a $\z_2 \times \z_2$ classification.  The first root state, referred to as the $\z_2$ root state, is obtained by placing the non-trivial $d=2$ SPT phase with $\z_2$ internal symmetry,\cite{chen2013cohomology, levin2012} which we refer to as the ``Ising SPT phase,'' on the mirror plane.  The second root state is obtained by placing an $E_8$ state\cite{kitaev11KITP} on the mirror plane.  While naively one might expect the $E_8$ root state to generate a $\z$ factor in the classification, this is not the case and it generates a $\z_2$ factor, as was shown in Ref.~\onlinecite{song2017}.

Each of these root states has a particularly simple surface termination, where the surface is gapped and trivial away from its intersection with the mirror plane.  The mirror axis on the surface is the gapless edge of the Ising SPT or $E_8$ state for the two root states, respectively.  Such simple surfaces will play an important role in the analysis of the present paper;  beginning with some $d=2$ field theory, we will add perturbations to make it trivial away from the mirror axis, and then study the resulting $d=1$ theory to match its anomaly with the Ising SPT or $E_8$ edge.

\begin{figure}
\includegraphics[width=0.9\columnwidth]{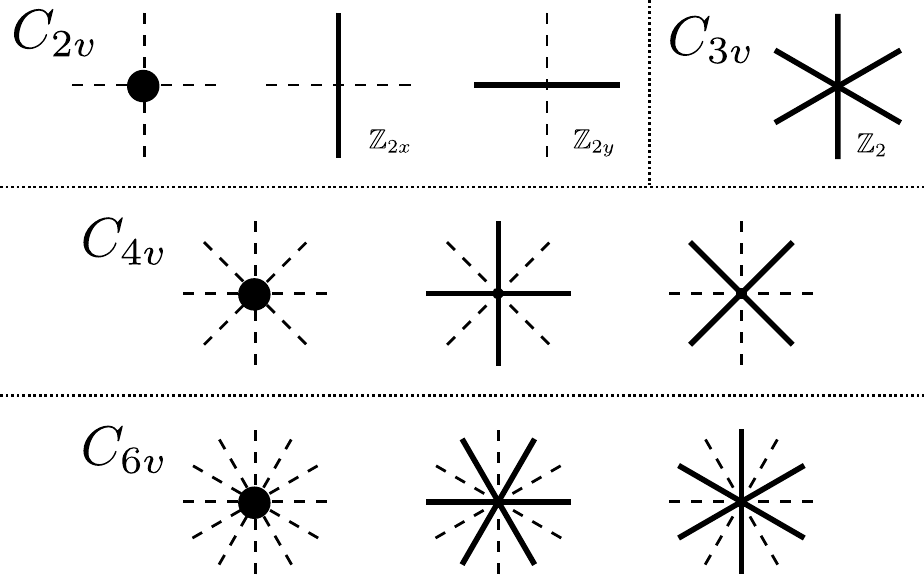}
\caption{Depiction of root states built from Haldane chains and Ising SPT states for $C_{nv}$ point groups.  In each case a cross-section normal to the $C_{nv}$ axis is shown.  Dashed lines are mirror planes hosting a trivial state, while solid lines represent Ising SPT states placed on mirror planes.  Solid circles represent Haldane chains placed on the $C_{nv}$ axis.}
\label{fig:rootstates}
\end{figure}

\begin{figure}
\includegraphics[width=0.9\columnwidth]{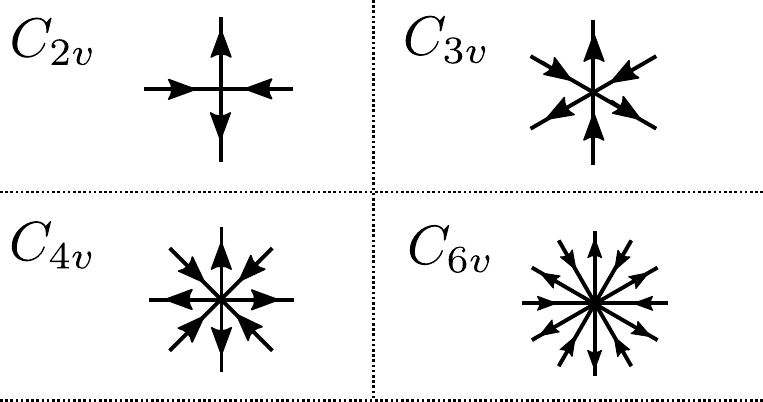}
\caption{Depiction of root states built from $E_{8}$ states for $C_{nv}$ point groups. Solid lines represent an $E_{8}$ state on each mirror plane. The arrows represent the edge chiralities of the $E_{8}$ states.}
\label{fig:rootstatesE8}
\end{figure}

A more interesting point group is $C_{2v}$, which is generated by two perpendicular mirror reflections, and where the lower-dimensional space is the union of the two mirror planes.  Here, there is a $\z_2^4$ classification of bosonic pgSPT phases, with four root states.\cite{song2017}  Two of these are obtained by placing an Ising SPT state on one mirror plane and leaving the other plane trivial; these root states are illustrated in Fig.~\ref{fig:rootstates}.  Another root state is obtained by placing $E_8$ states on the planes; up to equivalence operations, there is a unique arrangement that respects the $C_{2v}$ symmetry (see Fig.~\ref{fig:rootstatesE8}).  The last root state is obtained by placing a $d=1$ Haldane phase\cite{haldane1983, haldane1983nlsm, affleck1987} on the axis where the mirror planes intersect, on which $C_{2v}$ acts as an effective $\z_2 \times \z_2$ internal symmetry, which is sufficient to protect the Haldane phase.

In general, there are nine non-trivial $d=3$ point groups that can be preserved at a $d=2$ surface.  These are $C_{nv}$ and $C_n$, both with $n =2,3,4,6$, and $C_s$.  Only $C_s$ or $C_{nv}$ symmetry can protect non-trivial pgSPT phases, whose surface theories we discuss.   For $n \geq 3$, it is straightforward to apply the approach of Ref.~\onlinecite{song2017} to obtain a classification of pgSPT phases.  (This was obtained in Ref.~\onlinecite{huang17building} for the subset of phases built from lower-dimensional SPT states, ignoring phases built from $E_8$ states.)  For $n=3$, there is a single set of symmetry-related mirror planes, and the classification is $\z_2^2$.  One root state is obtained by placing Ising SPT states on the mirror planes, and the other by placing $E_8$ states.  For $n = 4,6$, there are two sets of symmetry-related mirror planes, and the classification is $\z_2^4$.  Two root states are obtained by placing Ising SPT states on different sets of mirror planes, one by placing $E_8$ states on the mirror planes, and one by placing a $d=1$ Haldane chain on the axis where the planes intersect, where the effective internal symmetry is $\z_n \rtimes \z_2$.  We note that, for $n=3$, $\z_3 \rtimes \z_2$ symmetry admits only trivial SPT phases in $d=1$; we do not obtain a non-trivial pgSPT phase by placing a Haldane chain on the symmetry axis for $C_{3v}$ symmetry.

The root states obtained by placing Ising SPT states on mirror planes and Haldane chains on $C_{nv}$ axes are illustrated in Fig.~\ref{fig:rootstates}.  $N_{f}=4$ QED3 is parent surface field for all these states, as discussed in Sec.~\ref{sec:4dirac}.  In Sec.~\ref{sec:e8}, we consider $\zz^P$ symmetry and show that QED3 with charge-1 and charge-3 fermions is a surface theory when an $E_8$ state is placed on the mirror plane.  Related root states for $C_{nv}$ symmetry, where $E_8$ states are placed on mirror planes, are shown Fig.~\ref{fig:rootstatesE8}.  We conjecture that QED3 with charge-1 and charge-3 fermions is also a parent surface field theory for these states.

Finally, we also consider $d=3$ bosonic SPT phases protected by ${\rm U}(1) \times \z_2^P$ symmetry, which we classify in Appendix~\ref{app:pgspt_u1m} using the dimensional reduction approach.  We find a $\z_2^4$ classification.  Two of the root states are obtained by placing an Ising SPT or an $E_8$ state on the mirror plane, and do not involve the ${\rm U}(1)$ symmetry.  Another root state is obtained by placing a boson integer quantum Hall (BIQH) state\cite{ymlu12theory,senthil2013} on the mirror plane.  The BIQH state is a $d=2$ SPT phase protected by ${\rm U}(1)$ symmetry.  The last root state is obtained by placing a two-dimensional ${\rm U}(1) \times \z_2$ SPT phase on the mirror plane that essentially involves both ${\rm U}(1)$ and $\z_2$ symmetries, in the sense that if either symmetry is broken the state becomes trivial.  A  parent surface field theory describing surfaces of all these phases except for the $E_8$ root state is discussed in Sec.~\ref{sec:nlsm}.

\subsection{Dirac theory of TCI surfaces}
\label{sec:dirac}

As a warm-up and a review, we consider the Dirac surface theory of TCIs with ${\rm U}(1) \times \z_{2}^{P}$ symmetry.  The non-interacting classification is $\z$, and it was shown by Isobe and Fu that interactions reduce this to $\z_8$.\cite{isobe15} The full classification is $\z_8 \times \z_2$, where the $\z_2$ factor is generated by the $E_8$ paramagnet TCI, which does not have a description in terms of non-interacting electrons.\cite{song2017}  This section essentially follows a part of the analysis of Ref.~\onlinecite{isobe15}, informed by the perspective of Ref.~\onlinecite{song2017}.

We focus on the root state generating the $\z_8$, which has a surface theory with a single Dirac fermion.  The $\z_8$ index of this state is $n = 1 \in \z_8$.  The same state can be obtained in dimensional reduction by placing a $\nu=1$ integer quantum Hall (IQH) state on the mirror plane, where we choose the mirror operation to act as the identity on the fermion fields.\cite{isobe15,song2017}

The surface theory is
\begin{equation}
\mathcal{L}_{\text{Dirac}}=-i\bar{\psi} \gamma_{\mu} \partial_{\mu} \psi,
\label{eqn:single_dirac}
\end{equation}
where $\gamma_{\mu}=\{ \tau^{z}, \tau^{x}, \tau^{y} \}$, $\bar{\psi}=\psi^{\dagger}(i \tau^{z})$, the $\tau^{\mu}$ are the $2 \times 2$ Pauli matrices, and sums over repeated indices are implied.  Very generally, reflection $M$ acts by
\begin{equation}
M : \psi(x,y) \to U_M \psi(-x, y) \text{,}
\end{equation}
where $U_M$ is a unitary matrix. Without loss of generality, we can take $U_M^2 = 1$, and it is straightforward to show there are  two solutions for $U_M$ that leave ${\cal L}_{\text{Dirac}}$ invariant, $U_M = \pm \tau^x = \pm \gamma_1$.  These solutions are the two distinct possibilities for the action of reflection symmetry on the continuum fields.  We will see in the following discussion that the two solutions result in surface theories for two different SPT phases.  Equivalently, we can say these solutions correspond to two different anomaly types of the surface theory.

We choose $U_M = \tau^x = \gamma_1$. To determine the anomaly type of this surface field theory, we dimensionally reduce it by adding a reflection preserving mass term. Passing to the Hamiltonian formalism, the Dirac Hamiltonian density is 
\begin{equation}
\mathcal{H}_{\text{Dirac}}=\psi^{\dagger} (i \tau^{y} \partial_{x} - i \tau^{x} \partial_{y}) \psi,
\label{}
\end{equation}
and we add the mass term:
\begin{equation}
\mathcal{H}_{m} = m(x) \psi^{\dagger} \tau^{z} \psi,
\label{}
\end{equation}
where $m(x)$ is a monotonic function with $m(x)=-m_{0}$ for $x \to -\infty$,  $m(x)=m_{0}$ for $x \to +\infty$, and $m_0$ a positive constant. We have thus introduced a single  domain wall where the mass changes sign, which is required by reflection symmetry.  The surface is gapped everywhere except on the $x=0$ mirror axis, and the anomaly type can be determined by studying any gapless modes pinned to the axis.

There is a single gapless mode with single-particle wave function
\begin{eqnarray}
| \psi_{p_{y}}(x,y) \rangle = \exp \left( i p_{y} y - \int_{0}^{x} m(x') dx' \right) | \chi_{+} \rangle,
\end{eqnarray}
where the spinor part is $| \chi_{+} \rangle = \frac{1}{\sqrt{2}} (1, 1)^{T}$.  The energy of this mode is $E(p_y) = p_y$, so the mode is chiral, and can be viewed as the edge of a $\nu = 1$ IQH state on the mirror plane.  Moreover, the mirror eigenvalue of this wave function is $\tau_M = 1$.  We note that taking $m_0 > 0$ was an arbitrary choice.  If instead  we take $m_0$ negative, we find a fermion mode of opposite chirality (therefore $\nu = -1$) and with $\tau_M = -1$.  The product $\tau_M \nu = 1$ remains unchanged, and can be associated with the $\z_8$ SPT invariant.

If we choose instead the solution $U_M = - \tau^x = - \gamma_1$,  and follow the same dimensional reduction procedure, we find a mode with the same chirality but mirror eigenvalue $\tau_{M}=-1$, corresponding to the surface of the $n = -1$ TCI. We see that different choices of symmetry action on the continuum fields indeed result in different anomaly types.

\section{QED3 parent field theory of bosonic point group SPT surfaces}
\label{sec:4dirac} 

Here, we consider quantum electrodynamics in three space-time dimensions (QED3) with $N_f = 4$ Dirac fermions as a parent surface theory of bosonic pgSPT phases.  First, in Sec.~\ref{sec:4dirac_tci} we consider mirror reflection symmetry.  We obtain $N_f = 4$ QED3 as a surface theory of the $\zz$ root state by gauging the surface of an electron TCI with four Dirac fermions.  We also discuss how the symmetry action can be modified to obtain a non-anomalous theory that can describe a purely two-dimensional system (or, equivalently, the surface of a trivial SPT phase).  We then proceed to consider other point groups, focusing primarily on pgSPT phases that are built by placing Ising SPT states on mirror planes.  In Sec.~\ref{sec:4dirac_c2v} we consider $C_{2v}$ symmetry.  We show that surfaces of the two root states built from Ising SPT states, and the state obtained by adding these two root states, can be described by $N_f = 4$ QED3 with different symmetry actions. In Sec.~\ref{sec:cnv}, we consider $C_{nv}$ symmetry for $n = 3,4,6$, showing that a set of root states built from Ising SPT states on mirror planes can all be described by the same surface field theory for appropriate actions of symmetry. Finally, in Sec.~\ref{sec:1d_haldane} we show that the $1d$ root state, obtained by placing a Haldane chain on the $C_{nv}$ axis for $n = 2,4,6$, also admits $N_f = 4$ QED3 as a surface theory.

\subsection{Mirror symmetry:  $\mathbb{Z}_{2}$ root state}
\label{sec:4dirac_tci}

For $d=3$ bosonic mirror SPT phases, the $\z_{2}$ root state is formed by placing a $d=2$ Ising SPT state on the mirror plane. It was shown in Ref.~\onlinecite{song2017} that the $n=4$ electronic TCI with $U(1) \times \z_{2}^{P}$ symmetry is equivalent to a product of the bosonic $\z_{2}$ root state with a trivial electronic insulator. We thus proceed to obtain a surface theory of the bosonic $\z_2$ root state by gauging the Dirac surface theory of the $n=4$ electronic TCI.

The surface of the $n=4$ electronic TCI can be described by a $N_f = 4$ Dirac theory: 
\begin{equation}
\mathcal{L}=-i\bar{\Psi} \gamma_{\mu} \partial_{\mu} \Psi,
\label{eqn:single_4dirac}
\end{equation}
where $\Psi$ is an $8$-component spinor comprised of $N_{f}=4$ Dirac fermions. We take tensor products of $\sigma^i$ and $\mu^i$ Pauli matrices ($i = x,y,z$) to act in the ${\rm SU}(4)$ flavor space, while $\tau^i$ Pauli matrices act in the two-component Dirac space as in Sec.~\ref{sec:dirac}.    We choose the reflection $x \to -x$ to act by 
\begin{equation}
M: \Psi(x,y) \rightarrow \tau^{x} \Psi(-x,y)=\gamma_{1} \Psi(-x,y). 
\label{eqn:m_action_40}
\end{equation}
We apply dimensional reduction by adding the mass term
\begin{equation}
\mathcal{L}_{m} = m(x) \bar{\Psi} \mu^{z} \Psi \text{,}
\end{equation}
with $m(x)$ as in Sec.~\ref{sec:dirac}.

Following Sec.~\ref{sec:dirac}, we find that there are two counter-propagating pairs of chiral fermions $c_I(y)$ on the mirror axis ($I = 1,\dots,4$). The effective $d=1$ Hamiltonian density is
\begin{equation}
\mathcal{H}_{1d} = -i v \sum_{I=1}^{2} c_{I}^{\dagger} \partial_{y} c_{I}+i v \sum_{I=3}^{4} c_{I}^{\dagger} \partial_{y} c_{I}, \label{eqn:h1d}
\end{equation}
with velocity $v$ and mirror symmetry action
\begin{eqnarray}
M: c_{1,2}(y) &\rightarrow& c_{1,2}(y) \nonumber \\
M: c_{3,4}(y) &\rightarrow& - c_{3,4}(y) \text{.} \label{eqn:mx-1d-action}
\end{eqnarray}
This theory was studied via bosonization in Refs.~\onlinecite{isobe15,song2017}.  By making an appropriate change of variables, Ref.~\onlinecite{song2017} showed the bosonized theory breaks into two decoupled sectors, where one sector is a trivial electronic insulator and the other is the edge of the Ising SPT state.  Therefore, the corresponding TCI is adiabatically connected to a product of a trivial insulator with the bosonic $\z_2$ root state.

To proceed, we first imagine the surface is in a limit where it is actually a product of a trivial electronic insulator and the edge of the Ising SPT state.  First of all, it is clear that we can consistently gauge the ${\rm U}(1)$ symmetry in the surface theory, because only degrees of freedom in the trivial insulating sector carry ${\rm U}(1)$ charge.  Second, gauging the ${\rm U}(1)$ symmetry using a compact gauge field turns the insulating sector into a trivial confined phase, leaving only the Ising SPT edge.\footnote{The gauge field must be compact so that it has a confined phase.  Equivalently, if the gauge field were non-compact, we would be introducing an undesired ${\rm U}(1)$ global symmetry associated with magnetic flux conservation.}  Therefore, gauging the ${\rm U}(1)$ produces a surface theory for the bosonic $\z_2$ root state.  These conclusions only have to do with anomalies of the surface, so they hold for any surface theory of the $n=4$ TCI.

We can thus obtain a surface theory of the bosonic $\z_2$ root state by starting with the Dirac theory Eq.~(\ref{eqn:single_4dirac}) and coupling the ${\rm U}(1)$ charge to a dynamical compact gauge field.  
The gauged Lagrangian is
\begin{equation}
\mathcal{L}_{g} = -i\bar{\Psi} \gamma^{\mu} (\partial_{\mu}+i a_{\mu}) \Psi+\mathcal{L}_{\text{Maxwell}} + \cdots
\label{eqn:gauge_4dirac}
\end{equation}
where $\mathcal{L}_{\text{Maxwell}}$ is the Maxwell term for $a_{\mu}$.  The ellipsis represents various perturbations consistent with microscopic symmetries, which include monopole operators because the gauge field is compact.  To keep ${\cal L}_g$ invariant under symmetry, we take $a_\mu$ even under reflection for $\mu = \tau,y$, with and $a_{x}$ odd.  This implies that electric charge is even under reflection, while magnetic flux is odd.

Qi and Fu have also obtained a $N_f = 4$ QED3 theory starting from the $n=4$ TCI surface, via a somewhat different route.\cite{qi2015anomalous}  Starting with the same Dirac surface theory, they introduced an emergent ${\rm U}(1)$ gauge field, and coupled this field to fermions to obtain a low-energy sector described by $N_f = 4$ QED3.  This theory takes a slightly different form from ours, but the two theories can be mapped to one another by a particle-hole transformation, as shown in Appendix~\ref{app:particle-hole}.  While these two QED3 theories are obtained by somewhat different routes, their equivalence is not a coincidence.  To obtain their QED3 theory, Qi and Fu integrate out high-energy excitations, which presumably corresponds to integrating out a trivial electronic insulating sector, thus producing a surface theory for the bosonic $\z_2$ root state.

We now describe how to modify the symmetry action in the $N_f = 4$ QED3 theory to obtain a surface theory of a trivial bosonic pgSPT surface.  This will be useful in our discussion of other crystalline symmetries below.  We modify the action of reflection symmetry by adding a Pauli matrix in the flavor space,
\begin{equation}
M : \Psi(x,y) \to \mu^x \tau^x \Psi(-x,y) = \mu^x \gamma_1 \Psi(x,y) \text{.}
\end{equation}
Such actions of reflection symmetry are known to occur in algebraic spin liquids of $d=2$ spin systems, \cite{hermele2005} where the effective theory and symmetry action is derived starting from a microscopic spin model using a parton effective field theory construction.  Therefore, we should expect this theory describes a trivial pgSPT surface.

Instead of appealing to the connection with a microscopic spin model, we show this QED3 theory describes a trivial pgSPT surface by adding perturbations to drive the theory into a trivial state. We can now add the spatially constant mass term ${\cal L}_m = m \bar{\Psi} \mu^z \Psi$, which gaps out the Dirac fermions everywhere.  It remains to consider the gauge sector, which, because the gauge field is compact and the matter fields are gapped, will be in a confining phase due to monopole proliferation and condensation.  We simply need to show it is possible for monopoles to condense without breaking the reflection symmetry.  Letting ${\cal M}$ be the unit charge monopole insertion operator, the most general possible action of reflection symmetry is $M : {\cal M} \to \lambda {\cal M}^\dagger$, where $\lambda$ is a phase factor, and the Hermitian conjugation appears because magnetic flux is odd under $M$.  It is possible to set $\lambda = 1$ by redefining the overall phase of ${\cal M}$, upon which the symmetry-invariant linear combination ${\cal M} + {\cal M}^\dagger$ can be added to the Lagrangian to confine the gauge sector without breaking reflection. 

\subsection{$C_{2v}$ symmetry}
\label{sec:4dirac_c2v}

Here, we discuss $N_f = 4$ QED3 as a surface theory of bosonic pgSPT phases protected by  $C_{2v}$ symmetry. The $C_{2v}$ point group is generated by two reflections, $M_{x}$ and $M_{y}$, whose mirror planes are perpendicular. The pgSPT classification is $(\mathbb{Z}_{2})^{4}$, in which two of the $\mathbb{Z}_{2}$ factors come from putting the $d=2$ Ising SPT state on either the $x=0$ or $y=0$ mirror plane.  We first discuss how to choose the action of $M_x$ and $M_y$ to realize these two $\z_2$ root states, which we refer to as the $\mathbb{Z}_{2x}$ and $\mathbb{Z}_{2y}$ root states, respectively.  Then we proceed to consider the surface of the $\mathbb{Z}_{2x} \oplus \mathbb{Z}_{2y}$ state obtained by adding the two root states together.

We consider the root state where an Ising SPT state is placed on the $x=0$ mirror plane, while the $y=0$ plane is trivial.  To obtain such a surface theory, we take
\begin{eqnarray}
M_{x}: \Psi(x,y) &\rightarrow& \gamma_{1} \Psi(-x,y), \nonumber
\\
M_{y}: \Psi(x,y) &\rightarrow& \gamma_{2} \mu^{x} \Psi(x,-y) \text{.}
\label{eqn:c2v_action}
\end{eqnarray}
This choice is motivated by the discussion of Sec.~\ref{sec:4dirac_tci}, where we showed that, for a single reflection symmetry, adding a flavor-space Pauli matrix to the reflection symmetry action results in a trivial pgSPT surface.  We can add the mass term
\begin{equation}
{\cal L}_m = m(x) \bar{\Psi} \mu^z \Psi \text{,}
\end{equation}
with $m(x)$ as above.  This gaps out the fermions away from $x=0$.  Moreover, we can repeat the discussion of Sec.~\ref{sec:4dirac_tci} to show that the monopoles can condense away from $x=0$ while preserving $M_y$ reflection symmetry.  Therefore, we have made the theory trivial away from the $x=0$ axis, as expected.

On the $x=0$ axis, ignoring coupling to the gauge field for the moment, we again obtain a $d=1$ theory of two counterpropagating pairs of chiral fermions.  The Hamiltonian density is the same as in Eq.~(\ref{eqn:h1d}), with $M_x$ acting as in Eq.~(\ref{eqn:mx-1d-action}) and 
\begin{eqnarray}
M_y : c_{1,2}(y) &\to& i c_{3,4}(-y) \text{,} \\
M_y : c_{3,4}(y) &\to& -i c_{1,2}(-y) \text{.}
\end{eqnarray}

To proceed, we need to show that it is possible to gauge the ${\rm U}(1)$ while preserving both $M_x$ and $M_y$.  This requires a generalization of the bosonization analysis of Ref.~\onlinecite{song2017} that was described above.  We introduce chiral boson fields $\phi_I$ by
\begin{eqnarray}
c^\dagger_{1,2} &\sim& e^{i \phi_{1,2}} \nonumber \\
c^\dagger_{3,4} &\sim& e^{-i \phi_{3,4}} \label{eqn:bosonization} \text{.}
\end{eqnarray}
If we were to simply repeat the analysis of Ref.~\onlinecite{song2017} at this point, we would make a change of variables to new fields $\phi'_I$, so that the fields $\phi'_{1,2}$ describe a trivial electronic insulator.  However, this cannot be done while maintaining both $M_x$ and $M_y$ symmetry.  The reason is that $M_x$ and $M_y$ anti-commute acting on fermion operators, which implies that in any theory of two counter-propagating fermion modes, one mode must have $M_x$ eigenvalue $1$, while the other mode has $M_x$ eigenvalue $-1$.  This is not a trivial electronic insulator, but rather is the edge of a non-trivial $d=1$ SPT phase protected by ${\rm U}(1)$ and $M_x$ symmetry.\cite{isobe15}

To circumvent this problem, we introduce two more chiral boson fields $\phi_{5,6}$.  We put these fields in a bosonic state that is trivial in the sense that it is the edge of a trivial SPT phase.  One is always free to add such trivial degrees of freedom at the edge; for instance, these fields can be interpreted physically as describing bulk degrees of freedom that we now include in the edge theory.  The bosonized theory is described by the Lagrangian
\begin{equation}
{\cal L} = \frac{1}{4\pi} K_{I J} \partial_t \phi_I \partial_y \phi_J  + \cdots \text{,}
\end{equation}
where the ellipsis represents local terms allowed by symmetry, which we will not need to consider explicitly.
The $K$ matrix is
\begin{equation}
K  = \left( \begin{array}{cccccc}
1 & 0 & 0 & 0 & 0 & 0 \\
0 & 1 & 0 & 0 & 0 & 0 \\
0 & 0 & -1 & 0 & 0 & 0 \\
0 & 0 & 0 & -1 & 0 & 0 \\
0 & 0 & 0 & 0 & 0 & 1 \\
0 & 0 & 0 & 0 & 1 & 0
\end{array}\right) \text{,}
\end{equation}
and symmetry acts on the bosonic fields by
\begin{eqnarray}
M_x : \vec{\phi} &\to& \vec{\phi} + \pi \vec{m}_x \\
M_y : \vec{\phi}(y) &\to& W_y \vec{\phi}(-y) + \pi \vec{m}_y \text{,} \\
{\rm U}(1) : \vec{\phi} &\to& \vec{\phi} + \alpha \vec{t} \text{,}
\end{eqnarray}
where $\vec{m}_x = (0,0,1,1,1,0)^T$, $\vec{m}_y = (-1/2,-1/2,-1/2,-1/2,0,0)^T$,
$\vec{t} =  (1,1,-1,-1,0,0)^T$, and
\begin{equation}
W_y = 
\left( \begin{array}{cccccc}
0 & 0 & -1 & 0 & 0 & 0 \\
0 & 0 & 0 & -1 & 0 & 0 \\
-1 & 0 & 0 & 0 & 0 & 0 \\
0 & -1 & 0 & 0 & 0 & 0 \\
0 & 0 & 0 & 0 & 1 & 0 \\
0 & 0 & 0 & 0 & 0 & -1
\end{array}\right) \text{.}
\end{equation}
We can confirm the bosonic $2 \times 2$ block is indeed trivial by noting that this sector can be gapped out by adding a $\cos(\phi_6)$ term, which respects all the symmetries, and which does not induce spontaneous symmetry breaking.

We change variables by
\begin{equation}
\phi' = W \phi \text{,}
\end{equation}
where $W$ is a ${\rm GL}(6, {\mathbb Z})$ matrix given by
\begin{equation}
W = 
\left( \begin{array}{cccccc}
1 & 0 & 0 & 0 & -1 & 0 \\
0 & 1 & 0 & 0 & 0 & 0 \\
0 & 0 & 1 & 0 & 1 & 0 \\
0 & 0 & 0 & 1 & 0 & 0 \\
0 & 0 & 0 & 0 & 1 & 0 \\
1 & 0 & 1 & 0 & 0 & 1
\end{array}\right) \text{.}
\end{equation}
The $K$ matrix is unaffected by this transformation, and the symmetries act by
\begin{eqnarray}
M_x : \vec{\phi}' &\to& \vec{\phi}' + \pi \vec{m}'_x \\
M_y : \vec{\phi}'(y) &\to& W'_y \vec{\phi}'(-y) + \pi \vec{m}'_y \\
{\rm U}(1) : \vec{\phi}' &\to& \vec{\phi}' + \alpha \vec{t'} \text{,}
\end{eqnarray}
where  $\vec{m}'_x = U \vec{m}_x = (-1,0,2,1,1,1)^T$, $\vec{m}'_y = U \vec{m}_y = (-1/2,-1/2,-1/2,-1/2,0,-1)$, $\vec{t'} =  U \vec{t} = \vec{t}$, and  $W'_y = W W_y W^{-1} = W_y$.

We can now refermionize the $\phi'_1, \dots, \phi'_4$ sector using the same transformation as in Eq.~(\ref{eqn:bosonization}), obtaining fermion fields $c'_I$.  We can gap out these fermions with the symmetry-preserving mass term
\begin{equation}
{\cal H}_m = m \Big[ (c'_1)^\dagger c'_4 - (c'_3)^\dagger c'_2 + \text{H.c.} \Big] \text{.}
\end{equation}
Therefore, this sector is a trivial electronic insulator.  The remaining $2 \times 2$ bosonic block is an Ising SPT edge under the $M_x$ symmetry.\cite{song2017}

We have thus shown that the $d=1$ theory decomposes into a trivial electronic insulator and an Ising SPT edge, so the corresponding TCI is a product of a trivial insulator with the bosonic root state obtained by placing an Ising SPT on the $x=0$ plane while keeping the $y = 0$ plane trivial.  Following the logic of  Sec.~\ref{sec:4dirac_tci}, we can thus gauge the ${\rm U}(1)$ while maintaining reflection symmetry to obtain a $N_f = 4$ QED3 surface theory for the ${\mathbb Z}_{2 x}$ root state, with symmetry action as in Eq.~(\ref{eqn:c2v_action}).  

Similarly, the symmetry action
\begin{eqnarray}
M_{x}: \Psi(x,y) &\rightarrow&  \mu^x \gamma_{1} \Psi(-x,y), \nonumber
\\
M_{y}: \Psi(x,y) &\rightarrow& \gamma_{2} \Psi(x,-y) \text{}
\label{eqn:c2v_action-z2y}
\end{eqnarray}
gives a surface theory for the $\mathbb{Z}_{2y}$ root state.

Now that we have obtained surface theories for the two root states, we consider the the $\mathbb{Z}_{2x} \oplus \mathbb{Z}_{2y}$ obtained by adding the two root states together.  One surface theory is simply two decoupled copies of $N_f = 4$ QED3, one with Dirac fermion field $\Psi_1$ on which $C_{2v}$ acts as in Eq.~(\ref{eqn:c2v_action}), and another with fermion $\Psi_2$ on which the symmetry action is given by Eq.~(\ref{eqn:c2v_action-z2y}).  From this starting point, we would like to obtain a single $N_f = 4$ QED3 surface theory.  We note that similar constructions will be used below to obtain surface theories for $C_{nv}$ pgSPT phases with $n \geq 3$.

To proceed, we introduce the 16 component field $\Upsilon = (\Psi_1, \Psi_2)^T$, and let $\varsigma^i$ Pauli matrices act in the additional two-dimensional flavor space.  We add the mass term
\begin{equation}
{\cal L}_m = m \bar{\Upsilon} X \Upsilon \text{,}  \label{eqn:c2v-coupling}
\end{equation}
where $X = \varsigma^x (1 - \mu^x)$.
This term is not fully gauge-invariant, and should be understood as the result of breaking the ${\rm U}(1) \times {\rm U}(1)$ gauge structure down to ${\rm U}(1)$ due to the Higgs mechanism.  We can diagonalize $X$ by making the change of variables $\Upsilon' = U \Upsilon$, where $U$ acts in the $\varsigma, \mu$ part of the flavor space and is given by
\begin{equation}
U = \frac{1}{2} \left( \begin{array}{cccc}
1 & -1 & -1 & 1 \\
-1 & 1 & -1 & 1 \\
0 & 0 & \sqrt{2} &  \sqrt{2} \\
 \sqrt{2} &  \sqrt{2} & 0 & 0
 \end{array}\right) \text{.}
 \end{equation}
 The mass matrix in the new basis becomes
 \begin{equation}
 X' = \left(\begin{array}{cccc}
 -2 & 0 & 0 & 0 \\
 0 & 2 & 0 & 0 \\
 0 & 0 & 0 & 0 \\
  0 & 0 & 0 & 0 \end{array}\right) \text{.}
\end{equation}

We integrate out the modes gapped out by the mass, and obtain a new $N_f = 4$ QED theory for the low-energy sector.  Denoting by $\Psi'$ the low-energy Dirac field, which is simply composed of those components of $\Upsilon'$ not gapped out, the symmetry action is
\begin{eqnarray}
M_x : \Psi' &\to& \gamma_1 \Psi' \\
M_y : \Psi' &\to& \gamma_2 \Psi' \text{.}
\end{eqnarray}

This form could have been guessed immediately, because upon dimensional reduction to either mirror plane one obtains an Ising SPT edge.  Indeed, another approach to show this is a correct surface theory would be to start from this $N_f = 4$ Dirac theory, carry out the dimensional reduction to the ``cross-shaped'' region containing both surface mirror axes, and then argue the ${\rm U}(1)$ symmetry can be gauged.  However, proceeding in this way one has to treat carefully the junction between the two $d=1$ theories at the $C_{2v}$ center, which seems likely to be more complicated than combining the two root state surface theories as we have done instead.

\subsection{$C_{nv}$ symmetry for $n=3,4,6$}
\label{sec:cnv}

The $C_{nv}$ point group is generated by a mirror reflection $M_y$ and $n$-fold rotation $C_n$, where the rotation axis lies within the mirror plane.  This results in $n$ mirror planes, as shown in Fig.~\ref{fig:rootstates}, where the root states obtained by placing Ising SPT states on mirror planes are shown.  For $n = 3$ there is a single such $\zz$ root state, while for $n = 4,6$ there are two root states labeled $\z_{2a}$ and $\z_{2b}$.

We first consider $C_{4v}$ symmetry.  We will obtain a surface field theory for the $\z_{2a}$ root state by stacking together two QED3 surface theories for $C_{2v}$ pgSPT phases, and adding a four-fold rotation symmetry that exchanges these two layers.  The resulting theory is two decoupled copies of $N_f = 4$ QED3, which nonetheless transform into one another under $C_{4v}$.  We then proceed along the same line as our discussion of the $C_{2v}$-symmetric $\z_{2x} \oplus \z_{2y}$ state, adding a mass term that couples the two layers and results in a single copy of $N_f = 4$ QED3.

We introduce eight-component Dirac fields $\Psi_1$ and $\Psi_2$ for the two layers, and combine these into the field $\Upsilon = (\Psi_1 \Psi_2)^T$.  Each of $\Psi_1$ and $\Psi_2$ is coupled to its own ${\rm U}(1)$ gauge field.  We choose symmetry to act by
\begin{eqnarray}
M_x : \Upsilon(x,y) &\to& \left(\begin{array}{cc} 
1 & 0 \\ 0 & \mu^x
\end{array}\right) \gamma_1 \Upsilon(-x,y) \\
M_y : \Upsilon(x,y) &\to& \left(\begin{array}{cc} 
\mu^x & 0 \\ 0 & 1
\end{array}\right) \gamma_2 \Upsilon(x,-y) \\
C_4 : \Upsilon(x,y) &\to& \left(\begin{array}{cc} 
0 & 1 \\ 1 & 0
\end{array}\right)  e^{i \pi \gamma_0 / 4} \Upsilon(-y, x) \text{,}
\end{eqnarray}
where the $2 \times 2$ matrices act in the additional two-dimensional flavor space (\emph{i.e.} the $\Psi_1$, $\Psi_2$ layer space).   It can be checked the relation $M_y = C_4 M_x C^{-1}_4$ is satisfied.

The same mass term as in Eq.~(\ref{eqn:c2v-coupling}) is allowed for these symmetries, and we can proceed as in that case to obtain a low-energy $N_f = 4$ QED3 theory with Dirac field $\Psi'$.  The action of symmetry on this field is found to be
\begin{eqnarray}
M_x : \Psi' &\to& \gamma_1 \Psi' \\
M_y : \Psi' &\to& \gamma_2 \Psi' \\
C_4 : \Psi' &\to& \mu^x e^{i \pi \gamma_0 / 4} \Psi' \text{.}
\end{eqnarray}
As expected, this symmetry action implies that if we reduce onto the $x=0$ or $y = 0$ mirror axis, we obtain an Ising SPT edge.  In addition, if we reduce onto the $x = y$ or $x = -y$ mirror axis, we obtain a trivial state.  To see this, we note that the $x \leftrightarrow y$ and $x \leftrightarrow -y$ reflections are $M_x C_4$ and $M_y C_4$, both of which act on $\Psi'$ with a $\mu^x$ Pauli matrix in the flavor space.  As shown in Sec.~\ref{sec:4dirac_tci}, this corresponds to the edge of a trivial state on these mirror planes.  Therefore, we have obtained a surface theory for the $\z_{2a}$ root state.  A similar theory for the $\z_{2b}$ root state can be obtained simply by rotating the coordinate axes.

For $C_{3v}$ symmetry, we introduce three $N_f = 4$ QED3 layers, each with field $\Psi_i$ ($i = 1,2,3$), and $\Upsilon = (\Psi_1 \Psi_2 \Psi_3)^T$.  We choose symmetry to act by
\begin{eqnarray}
M_x : \Upsilon &\to& \left( \begin{array}{ccc} 1 & 0 & 0 \\
0 & 0 & 1 \\
0 & 1 & 0
\end{array}\right) \gamma_1 \Upsilon \\
C_3 : \Upsilon &\to& \left( \begin{array}{ccc} 0 & 1 & 0 \\
0 & 0 & 1 \\
1 & 0 & 0
\end{array}\right) e^{i \pi \gamma_0 / 3} \Upsilon \text{,}
\end{eqnarray}
where the $3 \times 3$ matrices act in the layer space, and it can be checked that these transformations  satisfy the relations of the $C_{3v}$ group.  This choice is motivated by the fact that if we focus on $M_x$ and ignore the rest of the symmetry, then layer $i = 1$ hosts an Ising SPT edge on the mirror axis.  The same is true for $M_x C_3$ and layer $i=2$, and for $M_x C_3^2$ and layer $i=3$.  Therefore, we guess that stacking these three layers with the given symmetry action produces the $\z_2$ root state $C_{3v}$ pgSPT phase, as we now show.

We couple the layers by adding the $C_{3v}$ invariant term
\begin{equation}
{\cal L}_m = m \bar{\Upsilon} X \Upsilon \text{,}
\end{equation}
where
\begin{equation}
X = \left( \begin{array}{ccc}
0 & i & -i \\
-i & 0 & i \\
i & -i & 0
\end{array}\right) \text{.}
\end{equation}
This matrix has eigenvalues $0, \pm \sqrt{3}$, and it gaps out 8 of the 12 two-component fields, leaving the low-energy field
$\Psi' = (\Psi_1 + \Psi_2 + \Psi_3) / \sqrt{3}$.  Symmetry acts on this field by
\begin{eqnarray}
M_x : \Psi' &\to& \gamma_1 \Psi' \\
C_3 : \Psi' &\to& e^{i \pi \gamma_0 / 3} \Psi' \text{.}
\end{eqnarray}
In this theory, reducing onto any of the $M_x, M_x C_3, M_x C_3^2$ mirror axes results in the edge of an Ising SPT state, so we have indeed obtained a theory of the $\z_2$ root state $C_{3v}$ pgSPT surface.

Finally we consider $C_{6v}$ symmetry.  Again we introduce three $N_f = 4$ QED3 layers, with fields labeled as in the $C_{3v}$ case above.  Here, we take each layer to be invariant under the $C_{2v}$ subgroup generated by $M_x$ and $M_y$, while $C_6$ cyclically permutes the layers.  The symmetry action is
\begin{eqnarray}
M_x : \Upsilon &\to& \left( \begin{array}{ccc} 1 & 0 & 0 \\
0 & 0 & 1 \\
0 & 1 & 0
\end{array}\right) \gamma_1 \Upsilon \\
M_y : \Upsilon &\to& \left( \begin{array}{ccc} 1 & 0 & 0 \\
0 & 0 & 1 \\
0 & 1 & 0
\end{array}\right) \mu^x \gamma_2 \Upsilon \\
C_6 : \Upsilon &\to& \left( \begin{array}{ccc} 0 & 1 & 0 \\
0 & 0 & 1 \\
1 & 0 & 0
\end{array}\right)
e^{ i \pi \gamma_0 / 6} \mu^x \Upsilon \text{.}
\end{eqnarray}
This symmetry action allows the same mass term as in the $C_{3v}$ case above, which again leaves a low-energy field $\Psi'$, on which symmetry acts by
\begin{eqnarray}
M_x : \Psi' &\to& \gamma_1 \Psi' \\
M_y : \Psi' &\to& \mu^x \gamma_2 \Psi' \\
C_6 : \Psi' &\to& e^{i \pi \gamma_0 / 6} \mu^x \Psi' \text{.}
\end{eqnarray}
This is a $N_f = 4$ QED3 theory for the surface of the $\z_{2a}$ root state $C_{6v}$ pgSPT phase. Similarly, the surface theory for the $\z_{2b}$ root state can be obtained by a rotation.

\subsection{$1d$ root states : Haldane chain on the $C_{nv}$ axis}
\label{sec:1d_haldane}

While our main focus is on surfaces of pgSPT phases built from two-dimensional states, we briefly digress to discuss the $C_{nv}$ pgSPT phases built from a Haldane chain on the mirror axis, which occur for $n = 2,4,6$.\cite{huang17building}  We refer to this state as the $1d$ root state, and we show it also admits a $N_f = 4$ QED3 surface theory for an appropriate action of symmetry on the continuum fields.  We focus on $C_{2v}$, and briefly explain why similar results hold for $n = 4,6$ at the end of this section.

In the dimensional reduction picture, the surface of the $1d$ root state is a single projective representation of the effective $\zz \times \zz$ onsite symmetry, which lies at the $C_{2v}$ center where the surface mirror axes intersect.  In contrast to pgSPT phases built from two-dimensional states, this surface can occur in a strictly two-dimensional system, if spins transform projectively.  In that setting, the anomalous nature of the surface is manifest as a generalized Lieb-Schultz-Mattis constraint forbidding a trivial gapped state.  This bulk-boundary correspondence between Lieb-Schultz-Mattis constraints and certain SPT phases was proposed in Ref.~\onlinecite{cheng16translational}, and studied for $C_{2v}$ and other point group symmetries in Ref.~\onlinecite{huang17building}.  (See also Ref.~\onlinecite{yangqiLSM}.)

To obtain a surface theory of the $1d$ root state, we take advantage of existing works that start with a $S = 1/2$ Heisenberg model, and use parton gauge theory methods to obtain $N_f = 4$ QED3 as an effective low-energy theory.\cite{marston1989,wen2002,savary2017}  In this context, such states are often referred to as algebraic\cite{rantner2001} or Dirac\cite{ran2007} spin liquids.  For concreteness, we focus on the particular example of the $S = 1/2$ square lattice and the so-called staggered flux algebraic spin liquid, for which the  action of  microscopic symmetries on  continuum fields was obtained in Ref.~\onlinecite{hermele2005}.

Ref.~\onlinecite{hermele2005} considered square lattice symmetry, ${\rm SO}(3)$ spin rotation, and time reversal symmetry.  We consider the $C_{2v}$ subgroup generated by ``spin-orbit coupled reflections,'' which act on spin operators $\bS_{(x,y)}$ by
\begin{eqnarray}
M_x : \bS_{(x,y)} &\to& R_x(\pi) \bS_{(-x,y)} \\
M_y : \bS_{(x,y)} &\to& R_y(\pi) \bS_{(x,-y)} \text{,}
\end{eqnarray}
where $R_x(\pi)$ and $R_y(\pi)$ are spin rotations by the angle $\pi$ about $x$ and $y$ axes.  Such an action of symmetry is appropriate in a spin-orbit coupled system, and makes the spin at the origin into a projective representation of $C_{2v}$.

Using the results of Ref.~\onlinecite{hermele2005}, we obtain the following symmetry action on the continuum fermion field $\Psi$:
\begin{eqnarray}
M_x : \Psi(x,y) &\to& \sigma^x \mu^y \gamma_1 \Psi(-x,y) \\
M_y : \Psi(x,y) &\to& \sigma^y \mu^x \gamma_2 \Psi(x,-y) \text{.}
\end{eqnarray}
This defines a $N_f = 4$ QED3 surface theory of the $1d$ root state.

To obtain a similar surface field theory for the $1d$ root state with $C_{4v}$ symmetry, we can start with the same $S = 1/2$ algebraic spin liquid, and focus on a $C_{4v}$ subgroup made up of point group operations combined with appropriate spin rotations that make each $S = 1/2$ spin into a non-trivial projective representation of $C_{4v}$.  For $C_{6v}$ symmetry, we can consider the $\pi$-flux Dirac spin liquid on the $S = 1/2$ triangular lattice,\cite{ymlu16symmetric}, following the same strategy to choose an appropriate $C_{6v}$ subgroup.

\section{Cluston surface field theories of SPT phases built from $E_8$ states}
\label{sec:e8}

We now discuss surface field theories of bosonic and fermionic pgSPT phases built by placing an $E_8$ state on a mirror plane. In particular, we consider two related such SPT phases.  First, we consider the $E_8$ paramagnet TCI,\cite{song2017} which is a strongly interacting electronic SPT phase built by placing an $E_8$ state on the mirror plane.  Our surface field theory is based on the ``cluston'' construction of Ref.~\onlinecite{chong2015}, and is a non-interacting Dirac theory with both charge-1 and charge-3 Dirac fermions.  Once we obtain this surface field theory, we gauge the ${\rm U}(1)$ symmetry to obtain a theory for a bosonic pgSPT phase built by placing an $E_8$ state on the mirror plane. This theory is a variant of QED3 with both charge-1 and charge-3 Dirac fermions. We then discuss a route to obtain a gapped surface with three-fermion topological order from this field theory.

Electronic TCIs protected by $U(1) \times \z_2^P$ symmetry have a $\mathbb{Z}_{8} \times \mathbb{Z}_{2}$ classification,\cite{isobe15,song2017} where the $\mathbb{Z}_{2}$ factor is generated by the $E_{8}$ paramagnet TCI.\cite{song2017} In the dimensional reduction picture, the $E_{8}$ paramagnet TCI is described by placing a neutral bosonic $E_{8}$ state on the mirror plane, together with a trivial electronic insulator. The $E_8$ state can be characterized by its chiral central charge $c \operatorname{mod} 16=8$ and vanishing Hall conductance.  Upon coupling the electron charge to a compact ${\rm U}(1)$ gauge field, we eliminate the trivial fermionic sector and obtain a bosonic state with an $E_8$ state on the mirror plane.  This is a bosonic pgSPT state protected by $\z_2^P$, and more specifically is the $E_8$ root state of Ref.~\onlinecite{song2017}.  Because the fermionic sector is trivial, we can consistently gauge the ${\rm U}(1)$ symmetry on the surface alone; this gives a route to obtain surface theories of the $E_8$ root state.

An equivalent description of the $E_8$ paramagnet TCI can be obtained as follows. Starting with an $E_{8}$ state with $c=-8$ on the mirror plane, we also place a $\nu=8$ IQH state on the mirror plane.  This leaves the system in the same phase, because the $\nu=8$ IQH state alone on the mirror plane is a trivial pgSPT phase. The resulting state is characterized by $c=0$ and hall conductance $\nu=8$.  This is equivalent to placing a bosonic integer quantum Hall (BIQH) state  on the mirror plane, built from charge-2 Cooper pairs.  Therefore, the $E_8$ paramagnet TCI can equally well be viewed as the product of a BIQH state on the mirror plane with trivial gapped fermions.

We now give a Dirac surface theory of the $E_8$ paramagnet TCI, which we then justify using dimensional reduction.  The key insight comes from Ref.~\onlinecite{chong2015}, which showed that simple descriptions of the $d=2$ BIQH and $E_8$ states can be obtained by binding three electrons into ``cluston'' bound states, and then putting the clustons into a Chern band.  This results in a simple theory of non-interacting clustons and non-interacting (unbound) electrons.  In particular, the Cooper pair BIQH state can be obtained by combining a $\nu = 1$ IQH state of clustons with a $\nu = -1$ IQH state of electrons, resulting in $\nu = 8$ and central charge $c = 0$.  A description of a neutral $E_8$ state (embedded in an electron system) is obtained by combining nine copies of $\nu =1$ IQH states of electrons with a single copy of a $\nu = -1$ IQH state of clustons.

The surface theory has two Dirac fermion fields
\begin{eqnarray}
\mathcal{L}= -i \bar{\psi}  \gamma^{\mu} \partial_{\mu} \psi -i \bar{\psi}_{c}  \gamma^{\mu} \partial_{\mu} \psi_{c},
\label{eqn:cluston_dirac}
\end{eqnarray}
where $\psi$ carries charge-1 and $\psi_{c}$ is a charge-3 cluston. This was shown to be a surface theory of a strongly interacting topological insulator with $U(1) \rtimes \z_2^T$ symmetry, where $\z_2^T$ is time reversal.\cite{chong2015, seiberg2016}  To see that the same theory can describe the surface of the $E_8$ paramagnet TCI, we need to specify the mirror symmetry action, which we take to be
\begin{eqnarray}
M: \psi(x,y) &\rightarrow& -\tau_{x} \psi(-x,y) \text{,}  \label{eqn:mirror1}
\\
M: \psi_{c}(x,y) &\rightarrow& \tau_{x} \psi_{c}(-x,y) \text{.}  \label{eqn:mirror2}
\end{eqnarray}

To dimensionally reduce the surface, we add the following mass term to Eq.~\ref{eqn:cluston_dirac}:
\begin{equation}
\mathcal{L}_{m} = -m(x)  \bar{\psi} \psi +  m(x)  \bar{\psi}_{c} \psi_{c} \text{,}
\label{eqn:cluston_mass}
\end{equation}
with $m(x)$ as in Sec.~\ref{sec:dirac}.  This results in a counter-propagating pair of chiral modes on the mirror axis, described by the Hamiltonian density
\begin{eqnarray}
\mathcal{H}_{edge} &=& i v_{F} \psi^{\dagger} \partial_{y} \psi - i v_{F} \psi_{c}^{\dagger} \partial_{y} \psi_{c} \text{.}
\label{eqn:Hedge_cluston}
\end{eqnarray}
The symmetries act on the 1d fields by
\begin{eqnarray}
U(1): \psi^{\dagger} &\rightarrow& e^{i \alpha} \ \psi^{\dagger}\text{,}
\\
U(1): \psi_{c}^{\dagger} &\rightarrow& e^{i 3\alpha} \ \psi_{c}^{\dagger}\text{,}
\\
M: \psi^{\dagger} &\rightarrow&  \psi^{\dagger}\text{,}
\\
M: \psi_{c}^{\dagger} &\rightarrow& \psi_{c}^{\dagger} \text{.}
\end{eqnarray}
It follows immediately that the state on the mirror plane has $\nu = 8$ and $c = 0$, and is thus adiabatically connected to a Cooper pair BIQH state.  Therefore, Eq.~\ref{eqn:cluston_dirac} is a surface theory for the $E_8$ paramagnet TCI.

We can now gauge the $U(1)$ symmetry to obtain a surface field theory for the bosonic $E_{8}$ root state. The gauged Lagrangian has the following form:
\begin{eqnarray}
\mathcal{L}_{g} &=& -i \bar{\psi}  \gamma^{\mu} (\partial_{\mu}+i a_{\mu}) \psi -i \bar{\psi}_{c}  \gamma^{\mu} (\partial_{\mu}+ i 3 a_{\mu}) \psi_{c} \nonumber \\ &+& \mathcal{L}_{\text{Maxwell}} \text{.}
\label{eqn:cluston_dirac_gauged}
\end{eqnarray}
This field theory belongs to a series of self-dual theories studied in Ref.~\onlinecite{cheng2016}, labeled by the charge $k$ of $\psi_{c}$. In the large-$k$ limit the field theory flows to a CFT fixed point that can be studied in a $1/k$ expansion, and which may be relevant to the surface physics of pgSPT phases built from $E_8$ states, if it survives to $k=3$ and is sufficiently stable for realistic microscopic symmetries.

To connect this surface theory to the surface topological order, we introduce a charge-$2$ Higgs field $\phi$, which is odd under reflection, $M: \phi(x,y) \rightarrow -\phi(-x,y)$. 
 The Higgs field $\phi$ couples to the Dirac fermions in the following way:
\begin{eqnarray}
\mathcal{L}_{p}= i \phi^{*} \psi \tau_{y} \psi + i (\phi^{*})^{3} \psi_{c} \tau_{y} \psi_{c} + \text{H.c.}
\label{eqn:cluston_dirac_coupling}
\end{eqnarray}
This form of the coupling makes it clear that introducing $\phi$ is justified; the presence of the first term implies we can view $\phi$ as a charge-2 bound state of two $\psi$'s, which then couples to a bound state of two $\psi_c$'s as in the second term.

Condensing the Higgs field $\phi$ reduces the gauge structure down to $\mathbb{Z}_{2}$, resulting in a theory where a deconfined $\z_2$ gauge field is coupled to the fermions. In this Higgs phase, $\mathcal{L}_{p}$ becomes a pairing term,
\begin{eqnarray}
\mathcal{L}_{p}= i \Delta \psi \tau_{y} \psi + i \Delta^{3} \psi_{c} \tau_{y} \psi_{c} + \text{H.c.}
\label{eqn:cluston_dirac_pairing}
\end{eqnarray}
Here, $\Delta = \langle \phi \rangle$, and $\Delta$ and $-\Delta$ are two vacuua related by mirror symmetry.  These vacuua are gauge equivalent, so mirror symmetry is preserved in the Higgs phase.  It is thus convenient to redefine the mirror operation by combining it with a gauge transformation.  The new reflection symmetry $\tilde{M}$ acts on the fermions by
\begin{eqnarray}
\tilde{M}: \psi(x,y) &\rightarrow& -i \tau_{x} \psi(-x,y),
\label{eqn:mirror_gauge1}
\\
\tilde{M}: \psi_{c}(x,y) &\rightarrow& -i \tau_{x} \psi_{c}(-x,y).
\label{eqn:mirror_gauge2}
\end{eqnarray}

The properties of this surface superconductor have been analyzed in Ref.~\onlinecite{chong2015}. In particular, the fundamental vortex is a fermion, denoted by $e_{f}$ here. We identify the fermionic Bogoliubov quasi-particle as $\epsilon_{f}$. The two particles $e_{f}$ and $\epsilon_{f}$ have a non-trivial semionic braiding statistics since they see each other as a $\pi$-flux. The bound state of $e_{f}$ and $\epsilon_{f}$ is thus another fermionic anyon $m_{f}$. Therefore, the $\mathbb{Z}_{2}$ topological order we obtained is a reflection symmetric three-fermion state.

Now, we discuss the reflection symmetry fractionalization patterns of the three-fermion state, following Refs.~\onlinecite{yqi15detecting, zaletel15, song2017}.  Let $a$ be one of the three anyon types, and $S^a$ a string operator creating two $a$ particles in states that go into one another under reflection symmetry. Then, under reflection,
\begin{equation}
M : S^a \to \mu^a S^a \text{,}
\end{equation}
where $\mu^a = \pm 1$ characterizes the two possibilities for reflection symmetry fractionalization of $a$.  Because $\mu^{\epsilon_f} = \mu^{e_f} \mu^{m_f}$, it is enough to specify $\mu^a$ for $e_f$ and $m_f$.  Denoting $\mu^a = 1$ by 0 and $\mu^a = -1$ by $P$, the only two distinct fractionalization patterns are $e_f 0 m_f 0$ and $e_f P m_f P$.  Here we have taken into account that two fractionalization patterns related by a relabeling of anyons should be viewed as the same pattern.

It was shown in Ref.~\onlinecite{song2017} that the $E_8$ root state admits a surface with three-fermion topological order and the $e_f P m_f P$ reflection fractionalization pattern. From Eq.~\ref{eqn:mirror_gauge1} and~\ref{eqn:mirror_gauge2}, we know the reflection squares to $-1$ on a single Bogoliubov quasi-particle $\epsilon_{f}$.  Due to Fermi statistics, this implies $\mu^{\epsilon_f}  =1$,\cite{yqi15detecting,zaletel15} which is consistent with both fractionalization patterns.  The next question is then how does the reflection act on the fundamental vortex $e_{f}$. One way to obtain the reflection symmetry action on the vortex would be to solve the Bogolioubov de-Gennes equation for the vortex directly, and obtain the fractionalization pattern from that. Another way to proceed would be to observe that dimensional reduction tells us that we started with a surface theory for the $E_8$ root state, for which Ref.~\onlinecite{song2017} showed that the fractionalization pattern is $e_f P m_f P$.  The other $e_f 0 m_f 0$ fractionalization pattern is not possible, because it occurs at the surface of the $E_8 \oplus \z_2$ state. Therefore, the fractionalization pattern must be $e_f P m_f P$.

Here we present a different argument, based on an observation due to Chong Wang:\cite{chongwangprivate}  Recall that in  topological crystalline superconductors (TCSCs) with only reflection symmetry, the bosonic $E_{8}$ root state in the presence of trivial fermions is a trivial pgSPT phase,  no matter whether $M^{2}=\pm 1$ acting on electrons.\cite{song2017} Reinterpreting this result in terms of the surface topological order of the $E_{8}$ root state, it means that, in the presence of the electrons, the surface topological order can be trivialized by condensing a bosonic anyon, which is a bound state of an electron and one of the fermionic anyons of the three-fermion state. Applying this condition to the $e_{f} 0 m_{f} 0$ and $e_{f} P m_{f} P$ states, it is easy to see that only $e_{f} P m_{f} P$ can be trivialized in the presence of either $M^2 = + 1$ or $M^2 = -1$ electrons.  The other symmetry fractionalization pattern $e_{f} 0 m_{f} 0$ cannot be trivialized if $M^{2}=1$ acting on the electrons. This is because it corresponds to a product of the bosonic $E_{8} \oplus \mathbb{Z}_{2}$ root state with trivial fermions, which is a non-trivial TCSC.\cite{song2017}

We can use this observation to show the fractionalization pattern must be $e_{f} P m_{f} P$, by showing that the surface field theory also has the property that it can be trivialized in the presence of electrons on which $M^2 = 1$.  (There is no need to consider $M^2 = -1$ electrons.)  We start with the Lagrangian Eq.~\ref{eqn:cluston_dirac_gauged}, and introduce Dirac electrons $c_i$ with $i = 1, \dots, 8$ obeying the Lagrangian
\begin{equation}
{\cal L}_{{\text electron}} = -i\sum_{i = 1}^8 \bar{c}_i \gamma^{\mu} \partial_{\mu} c_i \text{,}
\end{equation}
and transforming under reflection by
\begin{equation}
M : c_i \to  \tau^x c_i \text{.}
\end{equation}
While our theory so far has a globally  conserved ${\rm U}(1)$ electron charge, we ignore this symmetry, so that the electron sector of the theory is the surface of a trivial TCSC.  We could equally well add terms breaking the global ${\rm U}(1)$, but there is no need to do this explicitly.  Now, we condense the Higgs field $c_8^\dagger \tau^z \psi$, which carries unit charge of the ${\rm U}(1)$ gauge field and thus eliminates the gauge field from the low-energy theory.  The low-energy theory thus has a term
\begin{equation}
{\cal L}_{\text{Higgs}} = m c_8^\dagger \tau^z \psi + \text{H.c.} \text{,}
\end{equation}
which gaps out both the $c_8$ and $\psi$ fermions.  This leaves us with the seven fermions $c_i$ ($i = 1,\dots,7$) and the single fermion $\psi_c$, all of which transform identically under mirror reflection, and none of which are coupled to a fluctuating gauge field.  Therefore, we have obtained a surface theory for a trivial TCSC.

\section{Bosonic parent surface field theory: non-linear sigma model and $CP^{1}$ model}
\label{sec:nlsm}

In this section, we use a coupled-wire approach to construct a surface theory for pgSPT phases with ${\rm U}(1) \times \zz^P$ symmetry.  We focus on states obtained by placing $d=2$ SPT states with ${\rm U}(1) \times \zz$ symmetry on the mirror plane, and the ``wires'' are the edges of a stack of such states at a symmetry-preserving surface. The field theory is an ${\rm O}(4)$ non-linear sigma model with a theta-term at $\theta = \pi$ and with ${\rm O}(2) \times {\rm O}(2)$ anisotropy, or, equivalently, a non-compact ${\rm CP}^1$ model.\cite{senthil2006}  The different pgSPT phases are distinguished by the symmetry action on the continuum fields.  Our approach is based on the construction of the ${\rm O}(4)$ sigma model in $d=2+1$ with theta term by coupling $d=1+1$ ${\rm SU}(2)_1$ Wess-Zumino-Witten (WZW) models.\cite{senthil2006}  Moreover, it is very closely related to, but conceptually somewhat simpler than, the network model constructions of surface field theories for bosonic topological insulators introduced in Ref.~\onlinecite{ashvin2013}.
  
In Appendix~\ref{app:pgspt_u1m}, we discuss the classification of bosonic pgSPT phases with ${\rm U}(1) \times \zz^P$ symmetry.  The classification is $\zz^4$, where one of the $\zz$ factors is generated by placing a neutral bosonic $E_{8}$ state on the mirror plane.  We will not consider the surface of this state in the following discussion. The remaining three  $\zz$ factors are generated by root states given by placing different ${\rm U}(1) \times \zz$ SPT states\cite{ymlu12theory, bi2015classification} on the mirror plane.  These states are the bosonic integer quantum Hall (BIQH) state, the Ising SPT state, and a state we refer to as the $\uzz$ state.  The $\uzz$ state is so-named because both symmetries are required to protect it.  Edges of these SPT states can be described in a chiral boson formulation,\cite{ymlu12theory} or equivalently as ${\rm SU}(2)_1$ WZW models.\cite{bi2015classification}  (We note that the $\zz^4$ classification ignores SPT states obtained by stacking BIQH layers of the same chirality; see Appendix~\ref{app:pgspt_u1m}.)

We follow the WZW model description, where the edge theory is
\begin{equation}
S=\int dy d\tau \frac{1}{2 \lambda} \text{tr} (\partial_{\mu}g^{\dagger}\partial_{\mu}g)+iS_{\text{WZW}}[g],
\label{eqn:wzw}
\end{equation}
where  $S_{\text{WZW}}[g]$ is the $SU(2)_{1}$ WZW term, and $g$  is a $SU(2)$ matrix field that we write as
\begin{equation}
g =
\left( \begin{array}{cc}
b_{1} & -b_{2}^{*} \\
b_{2} & b_{1}^{*}
\end{array}\right),
\end{equation}
with complex fields $b_1$ and $b_2$ satisfying $|b_1|^2 + |b_2|^2$.
We consider the following action of ${\rm U}(1)$ and $\zz$ symmetry:
\begin{eqnarray}
{\rm U}(1) : b_i &\to& e^{i n_i \alpha} b_i \label{eqn:u1z2p-action1} \\
M : b_i(y, \tau) : &\to& e^{i \pi m_i} b_i(y, \tau) \label{eqn:u1z2p-action2} \text{,}
\end{eqnarray}
where we write the generator of $\zz$ as $M$, anticipating its connection to $x \to -x$ mirror symmetry.  Here the four integers $n_i$ and $m_i$ are each either $0$ or $1$.  For the SPT edges of interest we have\cite{bi2015classification}
\begin{itemize}

\item BIQH state:  $n_1 = n_2 = 1$ and $m_1 = m_2 = 0$.  

\item Ising state:  $n_1 = n_2 = 0$, while $m_1 = m_2 = 1$.

\item $\uzz$ state:  $n_1 = m_2 = 1$ and $n_2 = m_1 = 0$.

\end{itemize}

We construct each of the root states by stacking the corresponding $d=2$ ${\rm U}(1) \times \zz$ SPT state on mirror planes.  We choose to stack along the $x$-direction, with $x = a J$ the $x$-coordinate of each plane, where $J$ is an integer and $a$ the lattice constant.  The discrete lattice symmetry of the stack is generated by $x \to -x$ reflection and translation $x \to x + 2 a$.  This choice of symmetry means that the $x = 0$ and $x = a$ mirror planes are not related by symmetry and can host different states.  This is important to obtain the BIQH root state, which is realized by stacking BIQH states of alternating chiralities.  The Ising and $\uzz$ root states are realized by simply stacking the corresponding $d=2$ SPT state on every mirror plane.

A symmetry-preserving surface is naturally described by coupling the edge states of each mirror plane.  The $x = a J$ edge is described by a ${\rm SU}(2)_1$ WZW model with field $g_J(y, \tau)$, and the action on the surface is
\begin{equation}
S_{\text{surface}}=S_{0}+S_{W}+S_{t},
\label{eqn:surface_biqh}
\end{equation}
where
\begin{eqnarray}
S_{0}&=&\sum_{J} \int dy d\tau \frac{1}{2 \lambda} \text{tr} (\partial_{\mu}g_{J}^{\dagger}\partial_{\mu}g_{J}),
\\
S_{W}&=& i \sum_{J} (-1)^{J} S_{WZW}[g_{J}],
\\
S_{t}&=& -t\sum_{J} \text{tr} (g^{\dagger}_{J}g_{J+1}+h.c.).
\end{eqnarray}
The action of ${\rm U}(1) \times \zz^P$ symmetry on $g_J$ follows from Eqs.~(\ref{eqn:u1z2p-action1}, \ref{eqn:u1z2p-action2}).

Though we are considering different symmetries, this is \emph{exactly} the same problem of coupled WZW models studied in Ref.~\onlinecite{senthil2006} and employed in Ref.~\onlinecite{ashvin2013}.  Ref.~\onlinecite{senthil2006} took the continuum limit and obtained the surface field theory
\begin{equation}
S_{\text{surface}}=\int d^{2}x d\tau \frac{1}{2 \kappa} \text{tr} (\partial_{\mu}g^{\dagger}\partial_{\mu}g)+i\pi S_{\theta},
\label{eqn:nlsm}
\end{equation} 
where 
\begin{equation}
S_{\theta} =  \int d^{2}x d\tau \frac{1}{24 \pi^{2}} \epsilon^{\mu \nu \lambda} \text{tr} \left( g^{-1}\partial_{\mu}g g^{-1}\partial_{\nu}g g^{-1}\partial_{\lambda}g \right).
\end{equation}
This is an ${\rm O}(4)$ non-linear sigma model with a topological theta-term (at $\theta=\pi$) as one can see by substituting $g=n_{0}+i \vec{n} \cdot \vec{\tau}$, for Pauli matrices $\tau^{\mu}$, and rewriting the action in terms of the ${\rm O}(4)$ vector $\hat{n}=(n_{0},\vec{n})$, where $n_{0}^{2}+(\vec{n})^{2}=1$.   The action of ${\rm U}(1)$ symmetry is still given by Eq.~(\ref{eqn:u1z2p-action1}), while reflection acts by
\begin{equation}
M :  b_i(x,y,\tau) \to e^{i \pi m_i} b_i(-x,y,\tau) \text{.}
\end{equation}

Ref.~\onlinecite{senthil2006} considered an ${\rm O}(2) \times {\rm O}(2)$ anisotropy, which corresponds to separate ${\rm U}(1)$ rotations of $b_1$ and $b_2$, and showed that this ${\rm O}(4)$ non-linear sigma model is dual to the easy-plane non-compact ${\rm CP}^1$ (NCCP$^1$) model, with action
\begin{equation}
S =  \sum_{s = \pm} \int d^2 x d\tau \,  | (\partial_{\mu} - i a_{1 \mu} ) \psi_{2 s} |^2 + S_{{\rm Maxwell}}[a_{1 \mu}] \text{.}
\end{equation}
Here we have a two-component complex field $\psi_{2 \pm}$ satisfying $|\psi_{2+}|^2 + |\psi_{2-}|^2 = 1$, and a non-compact ${\rm U}(1)$ vector potential $a_{1 \mu}$.  The second term is the usual Maxwell action for $a_{1 \mu}$.   We write $e^{i \chi_1} \sim b_1$ to denote the operator creating a unit-strength monopole in $a_{1 \mu}$. By following the duality transformation of Ref.~\onlinecite{senthil2006}, and keeping track of the  ${\rm U}(1) \times \zz^P$ symmetry, we obtain the symmetry action
\begin{eqnarray}
{\rm U}(1) : \Psi_2 &\to& \exp \Big( \frac{ i \alpha n_2}{2} \tau^z \Big) \Psi_2 \\
{\rm U}(1) : e^{i \chi_1} &\to&  e^{i n_1 \alpha} e^{i \chi_1} \\
M : \Psi_2 &\to& \tau^x \Psi_2^* \qquad \text{(}m_2 = 0\text{)} \\
M : \Psi_2 &\to& \tau^y \Psi_2^* \qquad \text{(}m_2 = 1\text{)} \\
M : e^{i \chi_1} &\to& e^{i \pi m_1} e^{i \chi_1} \text{,}
\end{eqnarray}
where $\Psi_2 = \left( \begin{array}{cc} \psi_{2+} & \psi_{2-} \end{array}\right)^T$.

Following Ref.~\onlinecite{ashvin2013} for time-reversal invariant bosonic topological insulators, one application of this surface field theory is to study gapped, topologically ordered surfaces.  This symmetry action allows us to condense the charge-2 field $\psi_{2+} \psi_{2-}$ without breaking symmetry, which results in a gapped $\zz$ gauge theory.  The field $\psi_{2+} \sim \psi^*_{2-}$ remains as a well-defined, gapped excitation carrying the $\zz$ gauge charge.  It carries half the ${\rm U}(1)$ charge if $n_2 = 1$.  If $m_2  = 1$, the $\zz$ gauge charge transforms with $M^2 = -1$, \emph{i.e.} it carries the non-trivial projective action of the reflection symmetry.\cite{essin2013classifying}

To identify the vison excitation, we follow Ref.~\onlinecite{ashvin2013} and take advantage of the self-duality of the easy-plane  NCCP$^1$ model.  Denoting the dual complex field by $\psi_{1\pm}$, and the dual gauge field and monopole creation operator by $a_{2 \mu}$ and $e^{i \chi_2}$, respectively, the symmetry action is found to be
\begin{eqnarray}
{\rm U}(1) : \Psi_1 &\to& \exp \Big( \frac{ i \alpha n_1}{2} \tau^z \Big) \Psi_1 \\
{\rm U}(1) : e^{i \chi_2} &\to&  e^{i n_2 \alpha} e^{i \chi_2} \\
M : \Psi_1 &\to& \tau^x \Psi_1^* \qquad \text{(}m_1 = 0\text{)} \\
M : \Psi_1 &\to& \tau^y \Psi_1^* \qquad \text{(}m_1 = 1\text{)} \\
M : e^{i \chi_2} &\to& e^{i \pi m_2} e^{i \chi_2} \text{.}
\end{eqnarray}
The transformation of $\psi_{1 \pm}$ implies that it carries $\pi$ flux of the $a_{1 \mu}$ gauge field, which identifies it as the vison in the topologically ordered phase.  Therefore, the vison carries half charge if $n_1 = 1$, and transforms with $M^2 = -1$ if $m_1 = 1$.

To summarize, depending on which state we are considering, we have the following fractionalization patterns:
\begin{itemize}

\item BIQH state:  $eCmC$.

\item Ising state:  $ePmP$.

\item $\uzz$ state:  $ePmC \simeq eCmP$.

\end{itemize}
Here, $e$ refers to the $\zz$ gauge charge and $m$ to the vison, while $C$ denotes half charge under ${\rm U}(1)$, and $P$ denotes $M^2 = -1$, \emph{i.e.} the non-trivial projective action of reflection symmetry.  Any two fractionalization patterns related by relabeling $e \leftrightarrow m$ are equivalent.  In previous work, Ref.~\onlinecite{hermele2016flux} identified $eCmP$ as an anomalous fractionalization pattern via a flux-fusion anomaly test; here, we identify the corresponding SPT bulk as the $\uzz$ state.  $ePmP$ and $eCmP$ are anomalous fractionalization patterns that can arise at a strongly interacting topological crystalline insulator surface,\cite{qi2015anomalous} and $ePmP$ was identified as a surface of the Ising state in Ref.~\onlinecite{song2017}.

It follows from these results that the pgSPT phase given by adding together the Ising and $\uzz$ states admits a gapped surface with $\zz$ topological order and $eCPmP$ fractionalization pattern, which was also identified as anomalous via the flux-fusion anomaly test.\cite{hermele2016flux}  To see this, we start with two decoupled, topologically ordered surface theories, with fractionalization patterns $e_1 P m_1 P$ and $e_2 C m_2 P$.  The composite $m_1 m_2$ has trivial symmetry fractionalization, so we condense it to obtain a surface with a single copy of $\zz$ topological order and the claimed fractionalization pattern.

\section{Discussion}
\label{sec:discussion}

In this paper, using dimensional reduction as a tool to reveal anomalies in a field theory, we considered three different surface field theories for bosonic pgSPT phases.  The first field theory is $N_{f}=4$ QED, obtained by gauging the $U(1)$ symmetry at the surface of a TCI with four Dirac fermions. Surfaces of all bosonic pgSPT phases built by placing $d=2$ Ising SPT states on mirror planes, or by placing Haldane chains on $C_{nv}$ axes, can be described by this theory.

The second field theory is a variant of QED3 with a single charge-1 and a single charge-3 Dirac fermions. This field theory is based on the "cluston" construction, introduced in Ref.~\onlinecite{chong2015} for various interacting topological phases involving time-reversal symmetry. We showed this field theory can describe the surface of the mirror-symmetric pgSPT phase built by placing an $E_{8}$ state on the mirror plane. We further discussed how to obtain from the field theory a gapped surface with three-fermion topological order and $e_{f}Pm_{f}P$ symmetry fractionalization.

The third field theory is the ${\rm O}(4)$ non-linear sigma model with a topological theta-term at $\theta = \pi$, which can be mapped to the ${\rm NCCP}^1$ model.\cite{senthil2006} Based on a coupled wire construction, we showed this field theory can describe the surfaces of three of the four root-state bosonic pgSPT phases with ${\rm U}(1) \times \zz^{P}$ symmetry.  These phases are built by  BIQH, Ising SPT, and ${\rm U}(1) - \zz$ states on the mirror plane.  Here also, we discussed the connection between the field theory and gapped, topologically ordered surfaces.

Although, for the latter two field theories, we focused on bosonic pgSPT phases with only one reflection symmetry, we conjecture that the same field theories can describe more general bosonic pgSPT phases involving multiple reflections.  For bosonic pgSPT phases built from $E_{8}$ states, the relevant point groups are $C_{nv}$ with $n=2, 3, 4, 6$, and Fig.~\ref{fig:rootstatesE8} shows the root states built by placing $E_{8}$ states on mirror planes.

If we start with the free Dirac theory with charge-1 and charge-3 fermions, and add a spatially varying mass term to implement dimensional reduction at the surface, we can obtain a pattern of BIQH states on mirror planes, where the edge chiralities match those shown in Fig.~\ref{fig:rootstatesE8}.  To proceed, it needs to be shown that the ${\rm U}(1)$ symmetry can be consistently gauged.  In the case of a single reflection symmetry, this is indeed the case, because the dimensionally reduced surface theory has the same anomaly type as a neutral $E_8$ state and trivial electronic insulator on the mirror plane.  For $C_{nv}$ symmetry, making a similar argument would require treating carefully the junction of the gapless edge states at the $C_{nv}$ axis.  We expect that such an analysis will lead to a clean separation of the theory into a neutral bosonic sector and a trivial fermionic sector carrying the ${\rm U}(1)$ charge, so that the ${\rm U}(1)$ symmetry can indeed be consistently gauged.

To obtain the ${\rm O}(4)$ non-linear sigma model with a theta-term at $\theta = \pi$ for bosonic pgSPT phases involving multiple reflection symmetries, one possibility would be to modify the coupled-wire construction to start with an array of wires invariant under the desired symmetry.  This would require treating couplings at junctions between wires in the process of taking the continuum limit.  An alternative route would be to start with decoupled copies of ${\rm O}(4)$ non-linear sigma models with a topological theta-term at $\theta = \pi$, that are taken to transform into one another under symmetry, similar to the approach followed for $N_f = 4$ QED3 theories in Sec.~\ref{sec:cnv}.  One would then need to add symmetry-allowed interactions with the goal of obtaining a single effective ${\rm O}(4)$ non-linear sigma model at $\theta = \pi$.  We leave such explicit demonstrations for future work.

\acknowledgments{S.-J.H. is grateful to Chong Wang and Liang Fu for useful discussions.  M.H. would like to thank T. Senthil for a useful discussion. We are both grateful to Liang Fu, Yi-Ping Huang and Hao Song for collaborations on related prior works. This work was supported by the U.S. Department of Energy, Office of Science, Basic Energy Sciences (BES) under Award number DE-SC0014415.}

\appendix
\section{Particle-hole transformation in $N_f = 4$ QED3}
\label{app:particle-hole}

In Ref.~\onlinecite{qi2015anomalous}, Qi and Fu introduced a field theory for the $n=4$ TCI surface. They also discussed how to obtain a gapped surface with $\zz$ topological order, with reflection squaring to minus one on $e$ and $m$ particles (the so-called $ePmP$ state) from the field theory they obtained. Here we show that Eq.~\ref{eqn:gauge_4dirac} can be mapped onto the field theory studied by Qi and Fu.

The mapping comprises a particle-hole transformation on some flavors of the fermions in Eq.~\ref{eqn:gauge_4dirac}. We first consider a single Dirac fermion in Eq.~\ref{eqn:single_dirac}. We define a particle-hole-transformed fermion 
\begin{eqnarray}
\psi' &=& \mathcal{C} \bar{\psi}^{T}=i \mathcal{C} \gamma^{0} \psi^{*}
\\
\bar{\psi'} &=& \psi'^{\dagger}(i\gamma^{0})=\psi'^{T}\gamma^{0} \mathcal{C}^{\dagger}\gamma^{0}.
\end{eqnarray}
The Lagrangian for the new fermion is
\begin{eqnarray}
\mathcal{L}'_{\text{Dirac}} &=& -i\bar{\psi'} \gamma^{\mu} \partial_{\mu} \psi'
\\
&=& -i (\psi^{T} \gamma^{0} \mathcal{C}^{\dagger} \gamma^{0}) \gamma^{\mu} (\mathcal{C} \partial_{\mu} \bar{\psi}^{T})
\\
&=& -i \bar{\psi} \mathcal{C}^{\dagger} (\gamma^{\mu})^{T} \gamma^{0} \mathcal{C}^{*} \gamma^{0} \partial_{\mu} \psi,
\end{eqnarray}
where in the last line we exchanged the two terms in the parentheses (giving a fermion minus sign) and integrated by parts. Putting $\mathcal{C}=\gamma^{2}$, we have 
\begin{eqnarray}
\mathcal{C}^{T} (\gamma^{\mu})^{T} \gamma^{0} \mathcal{C} \gamma^{0} &=& -\gamma^{2} \gamma^{\mu} \gamma^{2} 
\\
&=& \gamma^{\mu}.
\end{eqnarray}
Therefore, $\mathcal{L}_{\text{Dirac}}=\mathcal{L}'_{\text{Dirac}}$. 

Going back to the gauged theory Eq.~\ref{eqn:gauge_4dirac}, we perform the particle-hole transformation $\mathcal{C}$ to the third and fourth two-component Dirac fermions, and define 
\begin{eqnarray}
\Psi'=
\left( 
\begin{array}{c}
\psi_{1} \\
\psi_{2} \\
\psi_{3}' \\
\psi_{4}'
\end{array} 
\right ).
\end{eqnarray}
The transformed Lagrangian becomes 
\begin{equation}
\mathcal{L}'_{\text{g}} = -i\bar{\Psi}' \gamma^{\mu} (\partial_{\mu}+i \mu_{z} a_{\mu}) \Psi'+\mathcal{L}_{\text{Maxwell}}.
\label{eqn:gauge_4dirac_ph}
\end{equation}
Note that the action of reflection symmetry is unaltered by the transformation:
\begin{equation}
M: \Psi'(x,y) \rightarrow \gamma^{1} \Psi'(-x,y). 
\end{equation}
This is the same theory obtained by Qi and Fu.

\section{Classification of 3d bosonic pgSPT phases with $U(1) \times \z_{2}^{P}$ symmetry}
\label{app:pgspt_u1m}

Here we give the classification of $d=3$ bosonic pgSPT phases protected by $U(1) \times \z_{2}^{P}$ symmetry, using dimensional reduction. We first need the classification of invertible $d=2$ bosonic topological phases with on-site $U(1) \times \z_{2}$ symmetry.  The classification of 2d SPT phases with this symmetry was found to be $\z_2^2 \times \z$.\cite{chen2013cohomology, ymlu12theory}  In addition, since we are considering invertible topological phases and not just SPT phases, there is an additional $\z$ factor generated by an $E_8$ state, on whose degrees of freedom the symmetries act trivially.  Therefore the desired classification is $\z_2^2 \times \z^2$.

Actually, simply having the classification will not be enough for our purposes, as we need a more concrete description of the SPT states making up the $\zz^2 \times \z$.  This was obtained in Ref.~\onlinecite{ymlu12theory} by studying the edge states of these phases.
The edge of 2d bosonic SPT states is described by the Lagrangian 
\begin{eqnarray}
\mathcal{L}=\frac{1}{4\pi}(K_{ij}\partial_{x}\phi_{i}\partial_{t}\phi_{j}-V_{ij}\partial_{x}\phi_{i}\partial_{x}\phi_{j}),
\label{L}
\end{eqnarray}
with K-matrix
\begin{eqnarray}
K= \left( \begin{array}{cc}
0 & 1  \\
1 & 0  \\
\end{array} \right)
\end{eqnarray}
and $V$ is a $2 \times 2$ velocity matrix.
The symmetry acts  by
\begin{eqnarray}
U(1): \vec{\phi} &\rightarrow&  \vec{\phi}+ \theta \vec{t},
\end{eqnarray}
\begin{eqnarray}
M: \vec{\phi} &\rightarrow&  \vec{\phi}+\pi (m_{1},m_{2})^{T},
\end{eqnarray}
where $M$ is the generator of the $\zz$ symmetry, $\vec{t}_{e}=(1,q)^{T}$, with $q$ an arbitrary integer, and $m_1, m_2 = 0,1$. Different  SPT states are labeled by $[q,m_{1},m_{2}]$. The generator of the $\z$ factor is $[1,0,0]$, which is the bosonic integer quantum Hall (BIQH) state. \cite{ymlu12theory,senthil2013} The two $\zz$ factors are generated by $[0,1,1]$ and $[0,0,1]$.   The $[0,0,1]$ state is the $\uzz$ SPT phase discussed in Sec.~\ref{sec:nlsm}.  It can be shown that the $[0,1,1] \oplus [0,0,1]$ state is the same as the Ising SPT phase as discussed in Sec.~\ref{sec:nlsm}:  combining these two edge theories, it is possible to gap out modes to obtain a new theory with the same $2 \times 2$ $K$-matrix, where $\vec{t} = 0$ and $m_1 = m_2 = 1$.

To obtain the classification of $d=3$ pgSPT phases, we first need to apply dimensional reduction to obtain an invertible topological phase on the mirror plane.  In the present case, a pgSPT state can be trivialized away from the mirror plane as long as it is trivial ignoring the mirror symmetry and considering only the ${\rm U}(1)$ symmetry.  There are believed to be no $d=3$ SPT phases with only ${\rm U}(1)$ symmetry, so this will be the case.  Actually, there is an exception to this statement: a stack of BIQH states, all of the same chirality, is a non-trivial $d=3$ SPT phase.  In the absence of translation symmetry, it is subtle to characterize such a state; nonetheless, it has chiral charge transport at a ${\rm U}(1)$ symmetry-preserving surface, which is expected to be robust as long as ${\rm U}(1)$ symmetry is present.  For simplicity, we exclude such states from consideration, and focus only on states that become trivial when we consider only the ${\rm U}(1)$ symmetry.

Next, we need to consider the effect of adjoining layers.  It is possible to adjoin both $E_8$ and BIQH layers.  For the case with only $\zz^P$ symmetry, Ref.~\onlinecite{song2017} showed that adjoining $E_8$ layers reduces the $\z$ factor generated by the $E_8$ state in the $d=2$ classification to a $\zz$ factor in the $d=3$ classification.  The same conclusion holds here, because ${\rm U}(1)$ symmetry plays no role for these states.

It remains to understand the effect of adjoining BIQH layers. We will show that $[2,0,0]$ state is equivalent to a trivial state by adjoining $[-1,0,0]$ layers. We consider the state $[2,0,0] \oplus [-1,0,0] \oplus [-1,0,0]$, where the last two summands are the adjoined layers. The Lagrangian of the edge theory is of the same form as Eq.~(\ref{L}), with the $6 \times 6$ $K$-matrix
\begin{equation}
K  = \left( \begin{array}{cccccc}
0 & 1 & 0 & 0 & 0 & 0 \\
1 & 0 & 0 & 0 & 0 & 0 \\
0 & 0 & 0 & 1 & 0 & 0 \\
0 & 0 & 1 & 0 & 0 & 0 \\
0 & 0 & 0 & 0 & 0 & 1 \\
0 & 0 & 0 & 0 & 1 & 0
\end{array}\right) \text{.}
\end{equation}
Symmetry acts by
\begin{eqnarray}
M : \vec{\phi} &\to&  W_{M}\vec{\phi} \\
{\rm U}(1) : \vec{\phi} &\to& \vec{\phi} + \alpha \vec{t} \text{,}
\end{eqnarray}
where $\vec{t} = (1,2,1,-1,1,-1)^{T}$ and 
\begin{equation}
W_{M}  = \left( \begin{array}{cccccc}
1 & 0 & 0 & 0 & 0 & 0 \\
0 & 1 & 0 & 0 & 0 & 0 \\
0 & 0 & 0 & 0 & 1 & 0 \\
0 & 0 & 0 & 0 & 0 & 1 \\
0 & 0 & 1 & 0 & 0 & 0 \\
0 & 0 & 0 & 1 & 0 & 0
\end{array}\right) \text{.}
\end{equation}
Note that the mirror reflection symmetry exchanges the two adjoined layers, as is required.

Now we make a change of variables $\vec{\phi}' = W \vec{\phi}$, where $W$ is a ${\rm GL}(6, {\mathbb Z})$ matrix:
\begin{equation}
W  = \left( \begin{array}{cccccc}
1 & 0 & 0 & 0 & 0 & 0 \\
0 & 1 & 0 & 0 & 0 & 0 \\
0 & 0 & 1 & 0 & 0 & 0 \\
0 & 0 & 0 & 1 & 0 & 1 \\
0 & 0 & 0 & 0 & 0 & -1 \\
0 & 0 & 1 & 0 & -1 & 0
\end{array}\right) \text{.}
\end{equation}
The $K$-matrix is invariant under the change of variables. The $\vec{t}$ vector and the $W_{M}$ matrix in the new basis becomes $\vec{t'} = W\vec{t} = (1,2,1,-2,1,0)^{T}$ and 
\begin{equation}
W_{M}'  = \left( \begin{array}{cccccc}
1 & 0 & 0 & 0 & 0 & 0 \\
0 & 1 & 0 & 0 & 0 & 0 \\
0 & 0 & 1 & 0 & 0 & -1 \\
0 & 0 & 0 & 1 & 0 & 0 \\
0 & 0 & 0 & -1 & -1 & 0 \\
0 & 0 & 0 & 0 & 0 & -1
\end{array}\right) \text{.}
\end{equation}

All the edge modes can be gapped out while preserving symmetry, by the following two-step procedure. First of all, notice that $\vec{t'}=(1,2,1,-2,1,0)^{T}$ suggests that the lower-right $2\times 2$ block of the $K$-matrix describes a trivial bosonic state. We add a coupling $V_{1} = \lambda_{1}\cos{(\phi_{6}')}$, which pins $\phi_6' = 0$ and gaps out both $\phi_5'$ and $\phi_6'$. We integrate out the gapped modes to obtain a new $4 \times 4$ $K$-matrix, which consists of the upper-left $4 \times 4$ block of $K$,  where $\vec{t}' = \vec{t'}=(1,2,1,-2)^{T}$, and where $W'_M$ becomes the identity matrix.  This remaining sector looks like $[2,0,0] \oplus [-2,0,0]$. 

In the second step, we add the symmetry-preserving couplings $V_{2} = \lambda_{2}\cos{(\phi_{1}' - \phi_{3}')}$ and  $V_{3} = \lambda_{3}\cos{(\phi_{2}' + \phi_{4}')}$ to gap out all the remaining modes.  The arguments of these cosine terms commute, so the fields $\phi_{1}' - \phi_{3}'$ and $\phi_{2}' + \phi_{4}'$ can be pinned simultaneously.  The absence of spontaneous symmetry breaking can be checked by using the method introduced in Ref.~\onlinecite{levin2012kmatrix}.  We have thus gapped out the edge theory, and we conclude the $[2,0,0]$ state is equivalent to a trivial state upon adjoining BIQH layers.   Therefore, the $\z$ factor generated by BIQH state in the pgSPT classification reduces to a $\z_{2}$ factor.

\bibliography{sft}

\begin{thebibliography}{69}%
\makeatletter
\providecommand \@ifxundefined [1]{%
 \@ifx{#1\undefined}
}%
\providecommand \@ifnum [1]{%
 \ifnum #1\expandafter \@firstoftwo
 \else \expandafter \@secondoftwo
 \fi
}%
\providecommand \@ifx [1]{%
 \ifx #1\expandafter \@firstoftwo
 \else \expandafter \@secondoftwo
 \fi
}%
\providecommand \natexlab [1]{#1}%
\providecommand \enquote  [1]{``#1''}%
\providecommand \bibnamefont  [1]{#1}%
\providecommand \bibfnamefont [1]{#1}%
\providecommand \citenamefont [1]{#1}%
\providecommand \href@noop [0]{\@secondoftwo}%
\providecommand \href [0]{\begingroup \@sanitize@url \@href}%
\providecommand \@href[1]{\@@startlink{#1}\@@href}%
\providecommand \@@href[1]{\endgroup#1\@@endlink}%
\providecommand \@sanitize@url [0]{\catcode `\\12\catcode `\$12\catcode
  `\&12\catcode `\#12\catcode `\^12\catcode `\_12\catcode `\%12\relax}%
\providecommand \@@startlink[1]{}%
\providecommand \@@endlink[0]{}%
\providecommand \url  [0]{\begingroup\@sanitize@url \@url }%
\providecommand \@url [1]{\endgroup\@href {#1}{\urlprefix }}%
\providecommand \urlprefix  [0]{URL }%
\providecommand \Eprint [0]{\href }%
\providecommand \doibase [0]{http://dx.doi.org/}%
\providecommand \selectlanguage [0]{\@gobble}%
\providecommand \bibinfo  [0]{\@secondoftwo}%
\providecommand \bibfield  [0]{\@secondoftwo}%
\providecommand \translation [1]{[#1]}%
\providecommand \BibitemOpen [0]{}%
\providecommand \bibitemStop [0]{}%
\providecommand \bibitemNoStop [0]{.\EOS\space}%
\providecommand \EOS [0]{\spacefactor3000\relax}%
\providecommand \BibitemShut  [1]{\csname bibitem#1\endcsname}%
\let\auto@bib@innerbib\@empty
\bibitem [{\citenamefont {Kitaev}(2009)}]{kitaev2009}%
  \BibitemOpen
  \bibfield  {author} {\bibinfo {author} {\bibfnamefont {A.}~\bibnamefont
  {Kitaev}},\ }\href {\doibase 10.1063/1.3149495} {\bibfield  {journal}
  {\bibinfo  {journal} {AIP Conference Proceedings}\ }\textbf {\bibinfo
  {volume} {1134}},\ \bibinfo {pages} {22} (\bibinfo {year} {2009})},\ \Eprint
  {http://arxiv.org/abs/http://aip.scitation.org/doi/pdf/10.1063/1.3149495}
  {http://aip.scitation.org/doi/pdf/10.1063/1.3149495} \BibitemShut {NoStop}%
\bibitem [{\citenamefont {Ryu}\ \emph {et~al.}(2010)\citenamefont {Ryu},
  \citenamefont {Schnyder}, \citenamefont {Furusaki},\ and\ \citenamefont
  {Ludwig}}]{ryu2010}%
  \BibitemOpen
  \bibfield  {author} {\bibinfo {author} {\bibfnamefont {S.}~\bibnamefont
  {Ryu}}, \bibinfo {author} {\bibfnamefont {A.~P.}\ \bibnamefont {Schnyder}},
  \bibinfo {author} {\bibfnamefont {A.}~\bibnamefont {Furusaki}}, \ and\
  \bibinfo {author} {\bibfnamefont {A.~W.~W.}\ \bibnamefont {Ludwig}},\ }\href
  {http://stacks.iop.org/1367-2630/12/i=6/a=065010} {\bibfield  {journal}
  {\bibinfo  {journal} {New Journal of Physics}\ }\textbf {\bibinfo {volume}
  {12}},\ \bibinfo {pages} {065010} (\bibinfo {year} {2010})}\BibitemShut
  {NoStop}%
\bibitem [{\citenamefont {Pollmann}\ \emph {et~al.}(2010)\citenamefont
  {Pollmann}, \citenamefont {Turner}, \citenamefont {Berg},\ and\ \citenamefont
  {Oshikawa}}]{pollmann2010}%
  \BibitemOpen
  \bibfield  {author} {\bibinfo {author} {\bibfnamefont {F.}~\bibnamefont
  {Pollmann}}, \bibinfo {author} {\bibfnamefont {A.~M.}\ \bibnamefont
  {Turner}}, \bibinfo {author} {\bibfnamefont {E.}~\bibnamefont {Berg}}, \ and\
  \bibinfo {author} {\bibfnamefont {M.}~\bibnamefont {Oshikawa}},\ }\href
  {\doibase 10.1103/PhysRevB.81.064439} {\bibfield  {journal} {\bibinfo
  {journal} {Phys. Rev. B}\ }\textbf {\bibinfo {volume} {81}},\ \bibinfo
  {pages} {064439} (\bibinfo {year} {2010})}\BibitemShut {NoStop}%
\bibitem [{\citenamefont {Fidkowski}\ and\ \citenamefont
  {Kitaev}(2011)}]{fidkowski2011}%
  \BibitemOpen
  \bibfield  {author} {\bibinfo {author} {\bibfnamefont {L.}~\bibnamefont
  {Fidkowski}}\ and\ \bibinfo {author} {\bibfnamefont {A.}~\bibnamefont
  {Kitaev}},\ }\href {\doibase 10.1103/PhysRevB.83.075103} {\bibfield
  {journal} {\bibinfo  {journal} {Phys. Rev. B}\ }\textbf {\bibinfo {volume}
  {83}},\ \bibinfo {pages} {075103} (\bibinfo {year} {2011})}\BibitemShut
  {NoStop}%
\bibitem [{\citenamefont {Turner}\ \emph {et~al.}(2011)\citenamefont {Turner},
  \citenamefont {Pollmann},\ and\ \citenamefont {Berg}}]{turner2011}%
  \BibitemOpen
  \bibfield  {author} {\bibinfo {author} {\bibfnamefont {A.~M.}\ \bibnamefont
  {Turner}}, \bibinfo {author} {\bibfnamefont {F.}~\bibnamefont {Pollmann}}, \
  and\ \bibinfo {author} {\bibfnamefont {E.}~\bibnamefont {Berg}},\ }\href
  {\doibase 10.1103/PhysRevB.83.075102} {\bibfield  {journal} {\bibinfo
  {journal} {Phys. Rev. B}\ }\textbf {\bibinfo {volume} {83}},\ \bibinfo
  {pages} {075102} (\bibinfo {year} {2011})}\BibitemShut {NoStop}%
\bibitem [{\citenamefont {Chen}\ \emph
  {et~al.}(2011{\natexlab{a}})\citenamefont {Chen}, \citenamefont {Gu},\ and\
  \citenamefont {Wen}}]{chen2011_1dspt}%
  \BibitemOpen
  \bibfield  {author} {\bibinfo {author} {\bibfnamefont {X.}~\bibnamefont
  {Chen}}, \bibinfo {author} {\bibfnamefont {Z.-C.}\ \bibnamefont {Gu}}, \ and\
  \bibinfo {author} {\bibfnamefont {X.-G.}\ \bibnamefont {Wen}},\ }\href
  {\doibase 10.1103/PhysRevB.83.035107} {\bibfield  {journal} {\bibinfo
  {journal} {Phys. Rev. B}\ }\textbf {\bibinfo {volume} {83}},\ \bibinfo
  {pages} {035107} (\bibinfo {year} {2011}{\natexlab{a}})}\BibitemShut
  {NoStop}%
\bibitem [{\citenamefont {Chen}\ \emph
  {et~al.}(2011{\natexlab{b}})\citenamefont {Chen}, \citenamefont {Gu},\ and\
  \citenamefont {Wen}}]{chen2011_1dcomplete}%
  \BibitemOpen
  \bibfield  {author} {\bibinfo {author} {\bibfnamefont {X.}~\bibnamefont
  {Chen}}, \bibinfo {author} {\bibfnamefont {Z.-C.}\ \bibnamefont {Gu}}, \ and\
  \bibinfo {author} {\bibfnamefont {X.-G.}\ \bibnamefont {Wen}},\ }\href
  {\doibase 10.1103/PhysRevB.84.235128} {\bibfield  {journal} {\bibinfo
  {journal} {Phys. Rev. B}\ }\textbf {\bibinfo {volume} {84}},\ \bibinfo
  {pages} {235128} (\bibinfo {year} {2011}{\natexlab{b}})}\BibitemShut
  {NoStop}%
\bibitem [{\citenamefont {Schuch}\ \emph {et~al.}(2011)\citenamefont {Schuch},
  \citenamefont {P\'erez-Garc\'{\i}a},\ and\ \citenamefont
  {Cirac}}]{cirac2011}%
  \BibitemOpen
  \bibfield  {author} {\bibinfo {author} {\bibfnamefont {N.}~\bibnamefont
  {Schuch}}, \bibinfo {author} {\bibfnamefont {D.}~\bibnamefont
  {P\'erez-Garc\'{\i}a}}, \ and\ \bibinfo {author} {\bibfnamefont
  {I.}~\bibnamefont {Cirac}},\ }\href {\doibase 10.1103/PhysRevB.84.165139}
  {\bibfield  {journal} {\bibinfo  {journal} {Phys. Rev. B}\ }\textbf {\bibinfo
  {volume} {84}},\ \bibinfo {pages} {165139} (\bibinfo {year}
  {2011})}\BibitemShut {NoStop}%
\bibitem [{\citenamefont {Chen}\ \emph {et~al.}(2013)\citenamefont {Chen},
  \citenamefont {Gu}, \citenamefont {Liu},\ and\ \citenamefont
  {Wen}}]{chen2013cohomology}%
  \BibitemOpen
  \bibfield  {author} {\bibinfo {author} {\bibfnamefont {X.}~\bibnamefont
  {Chen}}, \bibinfo {author} {\bibfnamefont {Z.-C.}\ \bibnamefont {Gu}},
  \bibinfo {author} {\bibfnamefont {Z.-X.}\ \bibnamefont {Liu}}, \ and\
  \bibinfo {author} {\bibfnamefont {X.-G.}\ \bibnamefont {Wen}},\ }\href
  {\doibase 10.1103/PhysRevB.87.155114} {\bibfield  {journal} {\bibinfo
  {journal} {Phys. Rev. B}\ }\textbf {\bibinfo {volume} {87}},\ \bibinfo
  {pages} {155114} (\bibinfo {year} {2013})}\BibitemShut {NoStop}%
\bibitem [{\citenamefont {Levin}\ and\ \citenamefont {Gu}(2012)}]{levin2012}%
  \BibitemOpen
  \bibfield  {author} {\bibinfo {author} {\bibfnamefont {M.}~\bibnamefont
  {Levin}}\ and\ \bibinfo {author} {\bibfnamefont {Z.-C.}\ \bibnamefont {Gu}},\
  }\href {\doibase 10.1103/PhysRevB.86.115109} {\bibfield  {journal} {\bibinfo
  {journal} {Phys. Rev. B}\ }\textbf {\bibinfo {volume} {86}},\ \bibinfo
  {pages} {115109} (\bibinfo {year} {2012})}\BibitemShut {NoStop}%
\bibitem [{\citenamefont {Hasan}\ and\ \citenamefont
  {Kane}(2010)}]{hasan2010review}%
  \BibitemOpen
  \bibfield  {author} {\bibinfo {author} {\bibfnamefont {M.~Z.}\ \bibnamefont
  {Hasan}}\ and\ \bibinfo {author} {\bibfnamefont {C.~L.}\ \bibnamefont
  {Kane}},\ }\href {\doibase 10.1103/RevModPhys.82.3045} {\bibfield  {journal}
  {\bibinfo  {journal} {Rev. Mod. Phys.}\ }\textbf {\bibinfo {volume} {82}},\
  \bibinfo {pages} {3045} (\bibinfo {year} {2010})}\BibitemShut {NoStop}%
\bibitem [{\citenamefont {Qi}\ and\ \citenamefont
  {Zhang}(2011)}]{qi2011review}%
  \BibitemOpen
  \bibfield  {author} {\bibinfo {author} {\bibfnamefont {X.-L.}\ \bibnamefont
  {Qi}}\ and\ \bibinfo {author} {\bibfnamefont {S.-C.}\ \bibnamefont {Zhang}},\
  }\href {\doibase 10.1103/RevModPhys.83.1057} {\bibfield  {journal} {\bibinfo
  {journal} {Rev. Mod. Phys.}\ }\textbf {\bibinfo {volume} {83}},\ \bibinfo
  {pages} {1057} (\bibinfo {year} {2011})}\BibitemShut {NoStop}%
\bibitem [{\citenamefont {Hasan}\ and\ \citenamefont
  {Moore}(2011)}]{hasan2011review}%
  \BibitemOpen
  \bibfield  {author} {\bibinfo {author} {\bibfnamefont {M.~Z.}\ \bibnamefont
  {Hasan}}\ and\ \bibinfo {author} {\bibfnamefont {J.~E.}\ \bibnamefont
  {Moore}},\ }\href {\doibase 10.1146/annurev-conmatphys-062910-140432}
  {\bibfield  {journal} {\bibinfo  {journal} {Annual Review of Condensed Matter
  Physics}\ }\textbf {\bibinfo {volume} {2}},\ \bibinfo {pages} {55} (\bibinfo
  {year} {2011})},\ \Eprint
  {http://arxiv.org/abs/https://doi.org/10.1146/annurev-conmatphys-062910-140432}
  {https://doi.org/10.1146/annurev-conmatphys-062910-140432} \BibitemShut
  {NoStop}%
\bibitem [{\citenamefont {Senthil}(2015)}]{senthil2015review}%
  \BibitemOpen
  \bibfield  {author} {\bibinfo {author} {\bibfnamefont {T.}~\bibnamefont
  {Senthil}},\ }\href {\doibase 10.1146/annurev-conmatphys-031214-014740}
  {\bibfield  {journal} {\bibinfo  {journal} {Annual Review of Condensed Matter
  Physics}\ }\textbf {\bibinfo {volume} {6}},\ \bibinfo {pages} {299} (\bibinfo
  {year} {2015})},\ \Eprint
  {http://arxiv.org/abs/https://doi.org/10.1146/annurev-conmatphys-031214-014740}
  {https://doi.org/10.1146/annurev-conmatphys-031214-014740} \BibitemShut
  {NoStop}%
\bibitem [{\citenamefont {Chiu}\ \emph {et~al.}(2016)\citenamefont {Chiu},
  \citenamefont {Teo}, \citenamefont {Schnyder},\ and\ \citenamefont
  {Ryu}}]{chiu2016review}%
  \BibitemOpen
  \bibfield  {author} {\bibinfo {author} {\bibfnamefont {C.-K.}\ \bibnamefont
  {Chiu}}, \bibinfo {author} {\bibfnamefont {J.~C.~Y.}\ \bibnamefont {Teo}},
  \bibinfo {author} {\bibfnamefont {A.~P.}\ \bibnamefont {Schnyder}}, \ and\
  \bibinfo {author} {\bibfnamefont {S.}~\bibnamefont {Ryu}},\ }\href {\doibase
  10.1103/RevModPhys.88.035005} {\bibfield  {journal} {\bibinfo  {journal}
  {Rev. Mod. Phys.}\ }\textbf {\bibinfo {volume} {88}},\ \bibinfo {pages}
  {035005} (\bibinfo {year} {2016})}\BibitemShut {NoStop}%
\bibitem [{\citenamefont {Fu}(2011)}]{fu2011tci}%
  \BibitemOpen
  \bibfield  {author} {\bibinfo {author} {\bibfnamefont {L.}~\bibnamefont
  {Fu}},\ }\href {\doibase 10.1103/PhysRevLett.106.106802} {\bibfield
  {journal} {\bibinfo  {journal} {Phys. Rev. Lett.}\ }\textbf {\bibinfo
  {volume} {106}},\ \bibinfo {pages} {106802} (\bibinfo {year}
  {2011})}\BibitemShut {NoStop}%
\bibitem [{\citenamefont {Ando}\ and\ \citenamefont {Fu}(2015)}]{fu2015review}%
  \BibitemOpen
  \bibfield  {author} {\bibinfo {author} {\bibfnamefont {Y.}~\bibnamefont
  {Ando}}\ and\ \bibinfo {author} {\bibfnamefont {L.}~\bibnamefont {Fu}},\
  }\href {\doibase 10.1146/annurev-conmatphys-031214-014501} {\bibfield
  {journal} {\bibinfo  {journal} {Annual Review of Condensed Matter Physics}\
  }\textbf {\bibinfo {volume} {6}},\ \bibinfo {pages} {361} (\bibinfo {year}
  {2015})},\ \Eprint
  {http://arxiv.org/abs/https://doi.org/10.1146/annurev-conmatphys-031214-014501}
  {https://doi.org/10.1146/annurev-conmatphys-031214-014501} \BibitemShut
  {NoStop}%
\bibitem [{\citenamefont {Hsieh}\ \emph
  {et~al.}(2014{\natexlab{a}})\citenamefont {Hsieh}, \citenamefont {Sule},
  \citenamefont {Cho}, \citenamefont {Ryu},\ and\ \citenamefont
  {Leigh}}]{hsieh2014}%
  \BibitemOpen
  \bibfield  {author} {\bibinfo {author} {\bibfnamefont {C.-T.}\ \bibnamefont
  {Hsieh}}, \bibinfo {author} {\bibfnamefont {O.~M.}\ \bibnamefont {Sule}},
  \bibinfo {author} {\bibfnamefont {G.~Y.}\ \bibnamefont {Cho}}, \bibinfo
  {author} {\bibfnamefont {S.}~\bibnamefont {Ryu}}, \ and\ \bibinfo {author}
  {\bibfnamefont {R.~G.}\ \bibnamefont {Leigh}},\ }\href {\doibase
  10.1103/PhysRevB.90.165134} {\bibfield  {journal} {\bibinfo  {journal} {Phys.
  Rev. B}\ }\textbf {\bibinfo {volume} {90}},\ \bibinfo {pages} {165134}
  (\bibinfo {year} {2014}{\natexlab{a}})}\BibitemShut {NoStop}%
\bibitem [{\citenamefont {You}\ and\ \citenamefont {Xu}(2014)}]{you2014}%
  \BibitemOpen
  \bibfield  {author} {\bibinfo {author} {\bibfnamefont {Y.-Z.}\ \bibnamefont
  {You}}\ and\ \bibinfo {author} {\bibfnamefont {C.}~\bibnamefont {Xu}},\
  }\href {\doibase 10.1103/PhysRevB.90.245120} {\bibfield  {journal} {\bibinfo
  {journal} {Phys. Rev. B}\ }\textbf {\bibinfo {volume} {90}},\ \bibinfo
  {pages} {245120} (\bibinfo {year} {2014})}\BibitemShut {NoStop}%
\bibitem [{\citenamefont {Hsieh}\ \emph
  {et~al.}(2014{\natexlab{b}})\citenamefont {Hsieh}, \citenamefont {Morimoto},\
  and\ \citenamefont {Ryu}}]{hsieh2014cpt}%
  \BibitemOpen
  \bibfield  {author} {\bibinfo {author} {\bibfnamefont {C.-T.}\ \bibnamefont
  {Hsieh}}, \bibinfo {author} {\bibfnamefont {T.}~\bibnamefont {Morimoto}}, \
  and\ \bibinfo {author} {\bibfnamefont {S.}~\bibnamefont {Ryu}},\ }\href
  {\doibase 10.1103/PhysRevB.90.245111} {\bibfield  {journal} {\bibinfo
  {journal} {Phys. Rev. B}\ }\textbf {\bibinfo {volume} {90}},\ \bibinfo
  {pages} {245111} (\bibinfo {year} {2014}{\natexlab{b}})}\BibitemShut
  {NoStop}%
\bibitem [{\citenamefont {Isobe}\ and\ \citenamefont {Fu}(2015)}]{isobe15}%
  \BibitemOpen
  \bibfield  {author} {\bibinfo {author} {\bibfnamefont {H.}~\bibnamefont
  {Isobe}}\ and\ \bibinfo {author} {\bibfnamefont {L.}~\bibnamefont {Fu}},\
  }\href {\doibase 10.1103/PhysRevB.92.081304} {\bibfield  {journal} {\bibinfo
  {journal} {Phys. Rev. B}\ }\textbf {\bibinfo {volume} {92}},\ \bibinfo
  {pages} {081304} (\bibinfo {year} {2015})}\BibitemShut {NoStop}%
\bibitem [{\citenamefont {Cho}\ \emph {et~al.}(2015)\citenamefont {Cho},
  \citenamefont {Hsieh}, \citenamefont {Morimoto},\ and\ \citenamefont
  {Ryu}}]{cho2015}%
  \BibitemOpen
  \bibfield  {author} {\bibinfo {author} {\bibfnamefont {G.~Y.}\ \bibnamefont
  {Cho}}, \bibinfo {author} {\bibfnamefont {C.-T.}\ \bibnamefont {Hsieh}},
  \bibinfo {author} {\bibfnamefont {T.}~\bibnamefont {Morimoto}}, \ and\
  \bibinfo {author} {\bibfnamefont {S.}~\bibnamefont {Ryu}},\ }\href {\doibase
  10.1103/PhysRevB.91.195142} {\bibfield  {journal} {\bibinfo  {journal} {Phys.
  Rev. B}\ }\textbf {\bibinfo {volume} {91}},\ \bibinfo {pages} {195142}
  (\bibinfo {year} {2015})}\BibitemShut {NoStop}%
\bibitem [{\citenamefont {Yoshida}\ \emph {et~al.}(2015)\citenamefont
  {Yoshida}, \citenamefont {Morimoto},\ and\ \citenamefont
  {Furusaki}}]{yoshida2015}%
  \BibitemOpen
  \bibfield  {author} {\bibinfo {author} {\bibfnamefont {T.}~\bibnamefont
  {Yoshida}}, \bibinfo {author} {\bibfnamefont {T.}~\bibnamefont {Morimoto}}, \
  and\ \bibinfo {author} {\bibfnamefont {A.}~\bibnamefont {Furusaki}},\ }\href
  {\doibase 10.1103/PhysRevB.92.245122} {\bibfield  {journal} {\bibinfo
  {journal} {Phys. Rev. B}\ }\textbf {\bibinfo {volume} {92}},\ \bibinfo
  {pages} {245122} (\bibinfo {year} {2015})}\BibitemShut {NoStop}%
\bibitem [{\citenamefont {Kapustin}\ \emph {et~al.}(2015)\citenamefont
  {Kapustin}, \citenamefont {Thorngren}, \citenamefont {Turzillo},\ and\
  \citenamefont {Wang}}]{kapustin2015}%
  \BibitemOpen
  \bibfield  {author} {\bibinfo {author} {\bibfnamefont {A.}~\bibnamefont
  {Kapustin}}, \bibinfo {author} {\bibfnamefont {R.}~\bibnamefont {Thorngren}},
  \bibinfo {author} {\bibfnamefont {A.}~\bibnamefont {Turzillo}}, \ and\
  \bibinfo {author} {\bibfnamefont {Z.}~\bibnamefont {Wang}},\ }\href {\doibase
  10.1007/JHEP12(2015)052} {\bibfield  {journal} {\bibinfo  {journal} {Journal
  of High Energy Physics}\ }\textbf {\bibinfo {volume} {2015}},\ \bibinfo
  {pages} {1} (\bibinfo {year} {2015})}\BibitemShut {NoStop}%
\bibitem [{\citenamefont {Ware}\ \emph {et~al.}(2015)\citenamefont {Ware},
  \citenamefont {Kimchi}, \citenamefont {Parameswaran},\ and\ \citenamefont
  {Bauer}}]{sid2015}%
  \BibitemOpen
  \bibfield  {author} {\bibinfo {author} {\bibfnamefont {B.}~\bibnamefont
  {Ware}}, \bibinfo {author} {\bibfnamefont {I.}~\bibnamefont {Kimchi}},
  \bibinfo {author} {\bibfnamefont {S.~A.}\ \bibnamefont {Parameswaran}}, \
  and\ \bibinfo {author} {\bibfnamefont {B.}~\bibnamefont {Bauer}},\ }\href
  {\doibase 10.1103/PhysRevB.92.195105} {\bibfield  {journal} {\bibinfo
  {journal} {Phys. Rev. B}\ }\textbf {\bibinfo {volume} {92}},\ \bibinfo
  {pages} {195105} (\bibinfo {year} {2015})}\BibitemShut {NoStop}%
\bibitem [{\citenamefont {Qi}\ and\ \citenamefont
  {Fu}(2015{\natexlab{a}})}]{qi2015anomalous}%
  \BibitemOpen
  \bibfield  {author} {\bibinfo {author} {\bibfnamefont {Y.}~\bibnamefont
  {Qi}}\ and\ \bibinfo {author} {\bibfnamefont {L.}~\bibnamefont {Fu}},\ }\href
  {\doibase 10.1103/PhysRevLett.115.236801} {\bibfield  {journal} {\bibinfo
  {journal} {Phys. Rev. Lett.}\ }\textbf {\bibinfo {volume} {115}},\ \bibinfo
  {pages} {236801} (\bibinfo {year} {2015}{\natexlab{a}})}\BibitemShut
  {NoStop}%
\bibitem [{\citenamefont {Yoshida}\ and\ \citenamefont
  {Furusaki}(2015)}]{yoshida2015tci}%
  \BibitemOpen
  \bibfield  {author} {\bibinfo {author} {\bibfnamefont {T.}~\bibnamefont
  {Yoshida}}\ and\ \bibinfo {author} {\bibfnamefont {A.}~\bibnamefont
  {Furusaki}},\ }\href {\doibase 10.1103/PhysRevB.92.085114} {\bibfield
  {journal} {\bibinfo  {journal} {Phys. Rev. B}\ }\textbf {\bibinfo {volume}
  {92}},\ \bibinfo {pages} {085114} (\bibinfo {year} {2015})}\BibitemShut
  {NoStop}%
\bibitem [{\citenamefont {Morimoto}\ \emph {et~al.}(2015)\citenamefont
  {Morimoto}, \citenamefont {Furusaki},\ and\ \citenamefont
  {Mudry}}]{morimoto2015}%
  \BibitemOpen
  \bibfield  {author} {\bibinfo {author} {\bibfnamefont {T.}~\bibnamefont
  {Morimoto}}, \bibinfo {author} {\bibfnamefont {A.}~\bibnamefont {Furusaki}},
  \ and\ \bibinfo {author} {\bibfnamefont {C.}~\bibnamefont {Mudry}},\ }\href
  {\doibase 10.1103/PhysRevB.92.125104} {\bibfield  {journal} {\bibinfo
  {journal} {Phys. Rev. B}\ }\textbf {\bibinfo {volume} {92}},\ \bibinfo
  {pages} {125104} (\bibinfo {year} {2015})}\BibitemShut {NoStop}%
\bibitem [{\citenamefont {Fuji}\ \emph {et~al.}(2015)\citenamefont {Fuji},
  \citenamefont {Pollmann},\ and\ \citenamefont {Oshikawa}}]{fuji2015}%
  \BibitemOpen
  \bibfield  {author} {\bibinfo {author} {\bibfnamefont {Y.}~\bibnamefont
  {Fuji}}, \bibinfo {author} {\bibfnamefont {F.}~\bibnamefont {Pollmann}}, \
  and\ \bibinfo {author} {\bibfnamefont {M.}~\bibnamefont {Oshikawa}},\ }\href
  {\doibase 10.1103/PhysRevLett.114.177204} {\bibfield  {journal} {\bibinfo
  {journal} {Phys. Rev. Lett.}\ }\textbf {\bibinfo {volume} {114}},\ \bibinfo
  {pages} {177204} (\bibinfo {year} {2015})}\BibitemShut {NoStop}%
\bibitem [{\citenamefont {Lapa}\ \emph {et~al.}(2016)\citenamefont {Lapa},
  \citenamefont {Teo},\ and\ \citenamefont {Hughes}}]{lapa2016}%
  \BibitemOpen
  \bibfield  {author} {\bibinfo {author} {\bibfnamefont {M.~F.}\ \bibnamefont
  {Lapa}}, \bibinfo {author} {\bibfnamefont {J.~C.~Y.}\ \bibnamefont {Teo}}, \
  and\ \bibinfo {author} {\bibfnamefont {T.~L.}\ \bibnamefont {Hughes}},\
  }\href {\doibase 10.1103/PhysRevB.93.115131} {\bibfield  {journal} {\bibinfo
  {journal} {Phys. Rev. B}\ }\textbf {\bibinfo {volume} {93}},\ \bibinfo
  {pages} {115131} (\bibinfo {year} {2016})}\BibitemShut {NoStop}%
\bibitem [{\citenamefont {Hsieh}\ \emph {et~al.}(2016)\citenamefont {Hsieh},
  \citenamefont {Cho},\ and\ \citenamefont {Ryu}}]{hsieh2016}%
  \BibitemOpen
  \bibfield  {author} {\bibinfo {author} {\bibfnamefont {C.-T.}\ \bibnamefont
  {Hsieh}}, \bibinfo {author} {\bibfnamefont {G.~Y.}\ \bibnamefont {Cho}}, \
  and\ \bibinfo {author} {\bibfnamefont {S.}~\bibnamefont {Ryu}},\ }\href
  {\doibase 10.1103/PhysRevB.93.075135} {\bibfield  {journal} {\bibinfo
  {journal} {Phys. Rev. B}\ }\textbf {\bibinfo {volume} {93}},\ \bibinfo
  {pages} {075135} (\bibinfo {year} {2016})}\BibitemShut {NoStop}%
\bibitem [{\citenamefont {Lake}(2016)}]{lake2016}%
  \BibitemOpen
  \bibfield  {author} {\bibinfo {author} {\bibfnamefont {E.}~\bibnamefont
  {Lake}},\ }\href {\doibase 10.1103/PhysRevB.94.205149} {\bibfield  {journal}
  {\bibinfo  {journal} {Phys. Rev. B}\ }\textbf {\bibinfo {volume} {94}},\
  \bibinfo {pages} {205149} (\bibinfo {year} {2016})}\BibitemShut {NoStop}%
\bibitem [{\citenamefont {Song}\ and\ \citenamefont
  {Schnyder}(2017)}]{schnyder2017}%
  \BibitemOpen
  \bibfield  {author} {\bibinfo {author} {\bibfnamefont {X.-Y.}\ \bibnamefont
  {Song}}\ and\ \bibinfo {author} {\bibfnamefont {A.~P.}\ \bibnamefont
  {Schnyder}},\ }\href {\doibase 10.1103/PhysRevB.95.195108} {\bibfield
  {journal} {\bibinfo  {journal} {Phys. Rev. B}\ }\textbf {\bibinfo {volume}
  {95}},\ \bibinfo {pages} {195108} (\bibinfo {year} {2017})}\BibitemShut
  {NoStop}%
\bibitem [{\citenamefont {Song}\ \emph {et~al.}(2017)\citenamefont {Song},
  \citenamefont {Huang}, \citenamefont {Fu},\ and\ \citenamefont
  {Hermele}}]{song2017}%
  \BibitemOpen
  \bibfield  {author} {\bibinfo {author} {\bibfnamefont {H.}~\bibnamefont
  {Song}}, \bibinfo {author} {\bibfnamefont {S.-J.}\ \bibnamefont {Huang}},
  \bibinfo {author} {\bibfnamefont {L.}~\bibnamefont {Fu}}, \ and\ \bibinfo
  {author} {\bibfnamefont {M.}~\bibnamefont {Hermele}},\ }\href {\doibase
  10.1103/PhysRevX.7.011020} {\bibfield  {journal} {\bibinfo  {journal} {Phys.
  Rev. X}\ }\textbf {\bibinfo {volume} {7}},\ \bibinfo {pages} {011020}
  (\bibinfo {year} {2017})}\BibitemShut {NoStop}%
\bibitem [{\citenamefont {Jiang}\ and\ \citenamefont
  {Ran}(2017)}]{jiang17anyon}%
  \BibitemOpen
  \bibfield  {author} {\bibinfo {author} {\bibfnamefont {S.}~\bibnamefont
  {Jiang}}\ and\ \bibinfo {author} {\bibfnamefont {Y.}~\bibnamefont {Ran}},\
  }\href {\doibase 10.1103/PhysRevB.95.125107} {\bibfield  {journal} {\bibinfo
  {journal} {Phys. Rev. B}\ }\textbf {\bibinfo {volume} {95}},\ \bibinfo
  {pages} {125107} (\bibinfo {year} {2017})}\BibitemShut {NoStop}%
\bibitem [{\citenamefont {Thorngren}\ and\ \citenamefont
  {Else}(2016)}]{thorngren16gauging}%
  \BibitemOpen
  \bibfield  {author} {\bibinfo {author} {\bibfnamefont {R.}~\bibnamefont
  {Thorngren}}\ and\ \bibinfo {author} {\bibfnamefont {D.~V.}\ \bibnamefont
  {Else}},\ }\href@noop {} {\enquote {\bibinfo {title} {Gauging spatial
  symmetries and the classification of topological crystalline phases},}\ }
  (\bibinfo {year} {2016}),\ \Eprint {http://arxiv.org/abs/arXiv:1612.00846}
  {arXiv:1612.00846} \BibitemShut {NoStop}%
\bibitem [{\citenamefont {Huang}\ \emph {et~al.}(2017)\citenamefont {Huang},
  \citenamefont {Song}, \citenamefont {Huang},\ and\ \citenamefont
  {Hermele}}]{huang17building}%
  \BibitemOpen
  \bibfield  {author} {\bibinfo {author} {\bibfnamefont {S.-J.}\ \bibnamefont
  {Huang}}, \bibinfo {author} {\bibfnamefont {H.}~\bibnamefont {Song}},
  \bibinfo {author} {\bibfnamefont {Y.-P.}\ \bibnamefont {Huang}}, \ and\
  \bibinfo {author} {\bibfnamefont {M.}~\bibnamefont {Hermele}},\ }\href@noop
  {} {\enquote {\bibinfo {title} {Building crystalline topological phases from
  lower-dimensional states},}\ } (\bibinfo {year} {2017}),\ \Eprint
  {http://arxiv.org/abs/arXiv:1705.09243} {arXiv:1705.09243} \BibitemShut
  {NoStop}%
\bibitem [{\citenamefont {Fu}\ \emph {et~al.}(2007)\citenamefont {Fu},
  \citenamefont {Kane},\ and\ \citenamefont {Mele}}]{fu20073dti}%
  \BibitemOpen
  \bibfield  {author} {\bibinfo {author} {\bibfnamefont {L.}~\bibnamefont
  {Fu}}, \bibinfo {author} {\bibfnamefont {C.~L.}\ \bibnamefont {Kane}}, \ and\
  \bibinfo {author} {\bibfnamefont {E.~J.}\ \bibnamefont {Mele}},\ }\href
  {\doibase 10.1103/PhysRevLett.98.106803} {\bibfield  {journal} {\bibinfo
  {journal} {Phys. Rev. Lett.}\ }\textbf {\bibinfo {volume} {98}},\ \bibinfo
  {pages} {106803} (\bibinfo {year} {2007})}\BibitemShut {NoStop}%
\bibitem [{\citenamefont {Haldane}(1983{\natexlab{a}})}]{haldane1983}%
  \BibitemOpen
  \bibfield  {author} {\bibinfo {author} {\bibfnamefont {F.}~\bibnamefont
  {Haldane}},\ }\href {\doibase https://doi.org/10.1016/0375-9601(83)90631-X}
  {\bibfield  {journal} {\bibinfo  {journal} {Physics Letters A}\ }\textbf
  {\bibinfo {volume} {93}},\ \bibinfo {pages} {464 } (\bibinfo {year}
  {1983}{\natexlab{a}})}\BibitemShut {NoStop}%
\bibitem [{\citenamefont {Haldane}(1983{\natexlab{b}})}]{haldane1983nlsm}%
  \BibitemOpen
  \bibfield  {author} {\bibinfo {author} {\bibfnamefont {F.~D.~M.}\
  \bibnamefont {Haldane}},\ }\href {\doibase 10.1103/PhysRevLett.50.1153}
  {\bibfield  {journal} {\bibinfo  {journal} {Phys. Rev. Lett.}\ }\textbf
  {\bibinfo {volume} {50}},\ \bibinfo {pages} {1153} (\bibinfo {year}
  {1983}{\natexlab{b}})}\BibitemShut {NoStop}%
\bibitem [{\citenamefont {Affleck}\ \emph {et~al.}(1987)\citenamefont
  {Affleck}, \citenamefont {Kennedy}, \citenamefont {Lieb},\ and\ \citenamefont
  {Tasaki}}]{affleck1987}%
  \BibitemOpen
  \bibfield  {author} {\bibinfo {author} {\bibfnamefont {I.}~\bibnamefont
  {Affleck}}, \bibinfo {author} {\bibfnamefont {T.}~\bibnamefont {Kennedy}},
  \bibinfo {author} {\bibfnamefont {E.~H.}\ \bibnamefont {Lieb}}, \ and\
  \bibinfo {author} {\bibfnamefont {H.}~\bibnamefont {Tasaki}},\ }\href
  {\doibase 10.1103/PhysRevLett.59.799} {\bibfield  {journal} {\bibinfo
  {journal} {Phys. Rev. Lett.}\ }\textbf {\bibinfo {volume} {59}},\ \bibinfo
  {pages} {799} (\bibinfo {year} {1987})}\BibitemShut {NoStop}%
\bibitem [{\citenamefont {Wang}(2015)}]{chong2015}%
  \BibitemOpen
  \bibfield  {author} {\bibinfo {author} {\bibfnamefont {C.}~\bibnamefont
  {Wang}},\ }\href {\doibase 10.1103/PhysRevB.91.245124} {\bibfield  {journal}
  {\bibinfo  {journal} {Phys. Rev. B}\ }\textbf {\bibinfo {volume} {91}},\
  \bibinfo {pages} {245124} (\bibinfo {year} {2015})}\BibitemShut {NoStop}%
\bibitem [{\citenamefont {Seiberg}\ and\ \citenamefont
  {Witten}(2016)}]{seiberg2016}%
  \BibitemOpen
  \bibfield  {author} {\bibinfo {author} {\bibfnamefont {N.}~\bibnamefont
  {Seiberg}}\ and\ \bibinfo {author} {\bibfnamefont {E.}~\bibnamefont
  {Witten}},\ }\href {\doibase 10.1093/ptep/ptw083} {\bibfield  {journal}
  {\bibinfo  {journal} {Progress of Theoretical and Experimental Physics}\
  }\textbf {\bibinfo {volume} {2016}},\ \bibinfo {pages} {12C101} (\bibinfo
  {year} {2016})}\BibitemShut {NoStop}%
\bibitem [{\citenamefont {Senthil}\ and\ \citenamefont
  {Fisher}(2006)}]{senthil2006}%
  \BibitemOpen
  \bibfield  {author} {\bibinfo {author} {\bibfnamefont {T.}~\bibnamefont
  {Senthil}}\ and\ \bibinfo {author} {\bibfnamefont {M.~P.~A.}\ \bibnamefont
  {Fisher}},\ }\href {\doibase 10.1103/PhysRevB.74.064405} {\bibfield
  {journal} {\bibinfo  {journal} {Phys. Rev. B}\ }\textbf {\bibinfo {volume}
  {74}},\ \bibinfo {pages} {064405} (\bibinfo {year} {2006})}\BibitemShut
  {NoStop}%
\bibitem [{\citenamefont {Vishwanath}\ and\ \citenamefont
  {Senthil}(2013)}]{ashvin2013}%
  \BibitemOpen
  \bibfield  {author} {\bibinfo {author} {\bibfnamefont {A.}~\bibnamefont
  {Vishwanath}}\ and\ \bibinfo {author} {\bibfnamefont {T.}~\bibnamefont
  {Senthil}},\ }\href {\doibase 10.1103/PhysRevX.3.011016} {\bibfield
  {journal} {\bibinfo  {journal} {Phys. Rev. X}\ }\textbf {\bibinfo {volume}
  {3}},\ \bibinfo {pages} {011016} (\bibinfo {year} {2013})}\BibitemShut
  {NoStop}%
\bibitem [{\citenamefont {Wang}\ and\ \citenamefont
  {Senthil}(2014)}]{chong2014}%
  \BibitemOpen
  \bibfield  {author} {\bibinfo {author} {\bibfnamefont {C.}~\bibnamefont
  {Wang}}\ and\ \bibinfo {author} {\bibfnamefont {T.}~\bibnamefont {Senthil}},\
  }\href {\doibase 10.1103/PhysRevB.89.195124} {\bibfield  {journal} {\bibinfo
  {journal} {Phys. Rev. B}\ }\textbf {\bibinfo {volume} {89}},\ \bibinfo
  {pages} {195124} (\bibinfo {year} {2014})}\BibitemShut {NoStop}%
\bibitem [{\citenamefont {Witten}(2016)}]{witten2016}%
  \BibitemOpen
  \bibfield  {author} {\bibinfo {author} {\bibfnamefont {E.}~\bibnamefont
  {Witten}},\ }\href {\doibase 10.1103/RevModPhys.88.035001} {\bibfield
  {journal} {\bibinfo  {journal} {Rev. Mod. Phys.}\ }\textbf {\bibinfo {volume}
  {88}},\ \bibinfo {pages} {035001} (\bibinfo {year} {2016})}\BibitemShut
  {NoStop}%
\bibitem [{\citenamefont {You}\ \emph {et~al.}(2015)\citenamefont {You},
  \citenamefont {Bi}, \citenamefont {Rasmussen}, \citenamefont {Cheng},\ and\
  \citenamefont {Xu}}]{you2015}%
  \BibitemOpen
  \bibfield  {author} {\bibinfo {author} {\bibfnamefont {Y.-Z.}\ \bibnamefont
  {You}}, \bibinfo {author} {\bibfnamefont {Z.}~\bibnamefont {Bi}}, \bibinfo
  {author} {\bibfnamefont {A.}~\bibnamefont {Rasmussen}}, \bibinfo {author}
  {\bibfnamefont {M.}~\bibnamefont {Cheng}}, \ and\ \bibinfo {author}
  {\bibfnamefont {C.}~\bibnamefont {Xu}},\ }\href
  {http://stacks.iop.org/1367-2630/17/i=7/a=075010} {\bibfield  {journal}
  {\bibinfo  {journal} {New Journal of Physics}\ }\textbf {\bibinfo {volume}
  {17}},\ \bibinfo {pages} {075010} (\bibinfo {year} {2015})}\BibitemShut
  {NoStop}%
\bibitem [{\citenamefont {Bi}\ \emph {et~al.}(2015)\citenamefont {Bi},
  \citenamefont {Rasmussen}, \citenamefont {Slagle},\ and\ \citenamefont
  {Xu}}]{bi2015classification}%
  \BibitemOpen
  \bibfield  {author} {\bibinfo {author} {\bibfnamefont {Z.}~\bibnamefont
  {Bi}}, \bibinfo {author} {\bibfnamefont {A.}~\bibnamefont {Rasmussen}},
  \bibinfo {author} {\bibfnamefont {K.}~\bibnamefont {Slagle}}, \ and\ \bibinfo
  {author} {\bibfnamefont {C.}~\bibnamefont {Xu}},\ }\href {\doibase
  10.1103/PhysRevB.91.134404} {\bibfield  {journal} {\bibinfo  {journal} {Phys.
  Rev. B}\ }\textbf {\bibinfo {volume} {91}},\ \bibinfo {pages} {134404}
  (\bibinfo {year} {2015})}\BibitemShut {NoStop}%
\bibitem [{\citenamefont {Kitaev}()}]{kitaev11KITP}%
  \BibitemOpen
  \bibfield  {author} {\bibinfo {author} {\bibfnamefont {A.}~\bibnamefont
  {Kitaev}},\ }\href@noop {} {}\bibinfo {note} {\begin{scriptsize}
  {h}ttp://online.kitp.ucsb.edu/online/topomat11/kitaev\end{scriptsize}}\BibitemShut
  {NoStop}%
\bibitem [{\citenamefont {Lu}\ and\ \citenamefont
  {Vishwanath}(2012)}]{ymlu12theory}%
  \BibitemOpen
  \bibfield  {author} {\bibinfo {author} {\bibfnamefont {Y.-M.}\ \bibnamefont
  {Lu}}\ and\ \bibinfo {author} {\bibfnamefont {A.}~\bibnamefont
  {Vishwanath}},\ }\href {\doibase 10.1103/PhysRevB.86.125119} {\bibfield
  {journal} {\bibinfo  {journal} {Phys. Rev. B}\ }\textbf {\bibinfo {volume}
  {86}},\ \bibinfo {pages} {125119} (\bibinfo {year} {2012})}\BibitemShut
  {NoStop}%
\bibitem [{\citenamefont {Senthil}\ and\ \citenamefont
  {Levin}(2013)}]{senthil2013}%
  \BibitemOpen
  \bibfield  {author} {\bibinfo {author} {\bibfnamefont {T.}~\bibnamefont
  {Senthil}}\ and\ \bibinfo {author} {\bibfnamefont {M.}~\bibnamefont
  {Levin}},\ }\href {\doibase 10.1103/PhysRevLett.110.046801} {\bibfield
  {journal} {\bibinfo  {journal} {Phys. Rev. Lett.}\ }\textbf {\bibinfo
  {volume} {110}},\ \bibinfo {pages} {046801} (\bibinfo {year}
  {2013})}\BibitemShut {NoStop}%
\bibitem [{Note1()}]{Note1}%
  \BibitemOpen
  \bibinfo {note} {The gauge field must be compact so that it has a confined
  phase. Equivalently, if the gauge field were non-compact, we would be
  introducing an undesired ${\protect \rm U}(1)$ global symmetry associated
  with magnetic flux conservation.}\BibitemShut {Stop}%
\bibitem [{\citenamefont {Hermele}\ \emph {et~al.}(2005)\citenamefont
  {Hermele}, \citenamefont {Senthil},\ and\ \citenamefont
  {Fisher}}]{hermele2005}%
  \BibitemOpen
  \bibfield  {author} {\bibinfo {author} {\bibfnamefont {M.}~\bibnamefont
  {Hermele}}, \bibinfo {author} {\bibfnamefont {T.}~\bibnamefont {Senthil}}, \
  and\ \bibinfo {author} {\bibfnamefont {M.~P.~A.}\ \bibnamefont {Fisher}},\
  }\href {\doibase 10.1103/PhysRevB.72.104404} {\bibfield  {journal} {\bibinfo
  {journal} {Phys. Rev. B}\ }\textbf {\bibinfo {volume} {72}},\ \bibinfo
  {pages} {104404} (\bibinfo {year} {2005})}\BibitemShut {NoStop}%
\bibitem [{\citenamefont {Cheng}\ \emph {et~al.}(2016)\citenamefont {Cheng},
  \citenamefont {Zaletel}, \citenamefont {Barkeshli}, \citenamefont
  {Vishwanath},\ and\ \citenamefont {Bonderson}}]{cheng16translational}%
  \BibitemOpen
  \bibfield  {author} {\bibinfo {author} {\bibfnamefont {M.}~\bibnamefont
  {Cheng}}, \bibinfo {author} {\bibfnamefont {M.}~\bibnamefont {Zaletel}},
  \bibinfo {author} {\bibfnamefont {M.}~\bibnamefont {Barkeshli}}, \bibinfo
  {author} {\bibfnamefont {A.}~\bibnamefont {Vishwanath}}, \ and\ \bibinfo
  {author} {\bibfnamefont {P.}~\bibnamefont {Bonderson}},\ }\href {\doibase
  10.1103/PhysRevX.6.041068} {\bibfield  {journal} {\bibinfo  {journal} {Phys.
  Rev. X}\ }\textbf {\bibinfo {volume} {6}},\ \bibinfo {pages} {041068}
  (\bibinfo {year} {2016})}\BibitemShut {NoStop}%
\bibitem [{\citenamefont {Qi}\ \emph {et~al.}(2017)\citenamefont {Qi},
  \citenamefont {Fang},\ and\ \citenamefont {Fu}}]{yangqiLSM}%
  \BibitemOpen
  \bibfield  {author} {\bibinfo {author} {\bibfnamefont {Y.}~\bibnamefont
  {Qi}}, \bibinfo {author} {\bibfnamefont {C.}~\bibnamefont {Fang}}, \ and\
  \bibinfo {author} {\bibfnamefont {L.}~\bibnamefont {Fu}},\ }\href
  {https://arxiv.org/abs/1705.09190} {\enquote {\bibinfo {title} {Ground state
  degeneracy in quantum spin systems protected by crystal symmetries},}\ }
  (\bibinfo {year} {2017}),\ \Eprint {http://arxiv.org/abs/arXiv:1705.09190}
  {arXiv:1705.09190} \BibitemShut {NoStop}%
\bibitem [{\citenamefont {Marston}\ and\ \citenamefont
  {Affleck}(1989)}]{marston1989}%
  \BibitemOpen
  \bibfield  {author} {\bibinfo {author} {\bibfnamefont {J.~B.}\ \bibnamefont
  {Marston}}\ and\ \bibinfo {author} {\bibfnamefont {I.}~\bibnamefont
  {Affleck}},\ }\href {\doibase 10.1103/PhysRevB.39.11538} {\bibfield
  {journal} {\bibinfo  {journal} {Phys. Rev. B}\ }\textbf {\bibinfo {volume}
  {39}},\ \bibinfo {pages} {11538} (\bibinfo {year} {1989})}\BibitemShut
  {NoStop}%
\bibitem [{\citenamefont {Wen}(2002)}]{wen2002}%
  \BibitemOpen
  \bibfield  {author} {\bibinfo {author} {\bibfnamefont {X.-G.}\ \bibnamefont
  {Wen}},\ }\href {\doibase 10.1103/PhysRevB.65.165113} {\bibfield  {journal}
  {\bibinfo  {journal} {Phys. Rev. B}\ }\textbf {\bibinfo {volume} {65}},\
  \bibinfo {pages} {165113} (\bibinfo {year} {2002})}\BibitemShut {NoStop}%
\bibitem [{\citenamefont {Savary}\ and\ \citenamefont
  {Balents}(2017)}]{savary2017}%
  \BibitemOpen
  \bibfield  {author} {\bibinfo {author} {\bibfnamefont {L.}~\bibnamefont
  {Savary}}\ and\ \bibinfo {author} {\bibfnamefont {L.}~\bibnamefont
  {Balents}},\ }\href {http://stacks.iop.org/0034-4885/80/i=1/a=016502}
  {\bibfield  {journal} {\bibinfo  {journal} {Reports on Progress in Physics}\
  }\textbf {\bibinfo {volume} {80}},\ \bibinfo {pages} {016502} (\bibinfo
  {year} {2017})}\BibitemShut {NoStop}%
\bibitem [{\citenamefont {Rantner}\ and\ \citenamefont
  {Wen}(2001)}]{rantner2001}%
  \BibitemOpen
  \bibfield  {author} {\bibinfo {author} {\bibfnamefont {W.}~\bibnamefont
  {Rantner}}\ and\ \bibinfo {author} {\bibfnamefont {X.-G.}\ \bibnamefont
  {Wen}},\ }\href {\doibase 10.1103/PhysRevLett.86.3871} {\bibfield  {journal}
  {\bibinfo  {journal} {Phys. Rev. Lett.}\ }\textbf {\bibinfo {volume} {86}},\
  \bibinfo {pages} {3871} (\bibinfo {year} {2001})}\BibitemShut {NoStop}%
\bibitem [{\citenamefont {Ran}\ \emph {et~al.}(2007)\citenamefont {Ran},
  \citenamefont {Hermele}, \citenamefont {Lee},\ and\ \citenamefont
  {Wen}}]{ran2007}%
  \BibitemOpen
  \bibfield  {author} {\bibinfo {author} {\bibfnamefont {Y.}~\bibnamefont
  {Ran}}, \bibinfo {author} {\bibfnamefont {M.}~\bibnamefont {Hermele}},
  \bibinfo {author} {\bibfnamefont {P.~A.}\ \bibnamefont {Lee}}, \ and\
  \bibinfo {author} {\bibfnamefont {X.-G.}\ \bibnamefont {Wen}},\ }\href
  {\doibase 10.1103/PhysRevLett.98.117205} {\bibfield  {journal} {\bibinfo
  {journal} {Phys. Rev. Lett.}\ }\textbf {\bibinfo {volume} {98}},\ \bibinfo
  {pages} {117205} (\bibinfo {year} {2007})}\BibitemShut {NoStop}%
\bibitem [{\citenamefont {Lu}(2016)}]{ymlu16symmetric}%
  \BibitemOpen
  \bibfield  {author} {\bibinfo {author} {\bibfnamefont {Y.-M.}\ \bibnamefont
  {Lu}},\ }\href {\doibase 10.1103/PhysRevB.93.165113} {\bibfield  {journal}
  {\bibinfo  {journal} {Phys. Rev. B}\ }\textbf {\bibinfo {volume} {93}},\
  \bibinfo {pages} {165113} (\bibinfo {year} {2016})}\BibitemShut {NoStop}%
\bibitem [{\citenamefont {Cheng}\ and\ \citenamefont {Xu}(2016)}]{cheng2016}%
  \BibitemOpen
  \bibfield  {author} {\bibinfo {author} {\bibfnamefont {M.}~\bibnamefont
  {Cheng}}\ and\ \bibinfo {author} {\bibfnamefont {C.}~\bibnamefont {Xu}},\
  }\href {\doibase 10.1103/PhysRevB.94.214415} {\bibfield  {journal} {\bibinfo
  {journal} {Phys. Rev. B}\ }\textbf {\bibinfo {volume} {94}},\ \bibinfo
  {pages} {214415} (\bibinfo {year} {2016})}\BibitemShut {NoStop}%
\bibitem [{\citenamefont {Qi}\ and\ \citenamefont
  {Fu}(2015{\natexlab{b}})}]{yqi15detecting}%
  \BibitemOpen
  \bibfield  {author} {\bibinfo {author} {\bibfnamefont {Y.}~\bibnamefont
  {Qi}}\ and\ \bibinfo {author} {\bibfnamefont {L.}~\bibnamefont {Fu}},\ }\href
  {\doibase 10.1103/PhysRevB.91.100401} {\bibfield  {journal} {\bibinfo
  {journal} {Phys. Rev. B}\ }\textbf {\bibinfo {volume} {91}},\ \bibinfo
  {pages} {100401} (\bibinfo {year} {2015}{\natexlab{b}})}\BibitemShut
  {NoStop}%
\bibitem [{\citenamefont {Zaletel}\ \emph {et~al.}(2015)\citenamefont
  {Zaletel}, \citenamefont {Lu},\ and\ \citenamefont {Vishwanath}}]{zaletel15}%
  \BibitemOpen
  \bibfield  {author} {\bibinfo {author} {\bibfnamefont {M.}~\bibnamefont
  {Zaletel}}, \bibinfo {author} {\bibfnamefont {Y.-M.}\ \bibnamefont {Lu}}, \
  and\ \bibinfo {author} {\bibfnamefont {A.}~\bibnamefont {Vishwanath}},\
  }\href@noop {} {\enquote {\bibinfo {title} {Measuring space-group symmetry
  fractionalization in z$_2$ spin liquids},}\ } (\bibinfo {year} {2015}),\
  \Eprint {http://arxiv.org/abs/arXiv:1501.01395} {arXiv:1501.01395}
  \BibitemShut {NoStop}%
\bibitem [{\citenamefont {Wang}()}]{chongwangprivate}%
  \BibitemOpen
  \bibfield  {author} {\bibinfo {author} {\bibfnamefont {C.}~\bibnamefont
  {Wang}},\ }\href@noop {} {}\bibinfo {note} {{p}rivate
  communication}\BibitemShut {NoStop}%
\bibitem [{\citenamefont {Essin}\ and\ \citenamefont
  {Hermele}(2013)}]{essin2013classifying}%
  \BibitemOpen
  \bibfield  {author} {\bibinfo {author} {\bibfnamefont {A.~M.}\ \bibnamefont
  {Essin}}\ and\ \bibinfo {author} {\bibfnamefont {M.}~\bibnamefont
  {Hermele}},\ }\href {\doibase 10.1103/PhysRevB.87.104406} {\bibfield
  {journal} {\bibinfo  {journal} {Phys. Rev. B}\ }\textbf {\bibinfo {volume}
  {87}},\ \bibinfo {pages} {104406} (\bibinfo {year} {2013})}\BibitemShut
  {NoStop}%
\bibitem [{\citenamefont {Hermele}\ and\ \citenamefont
  {Chen}(2016)}]{hermele2016flux}%
  \BibitemOpen
  \bibfield  {author} {\bibinfo {author} {\bibfnamefont {M.}~\bibnamefont
  {Hermele}}\ and\ \bibinfo {author} {\bibfnamefont {X.}~\bibnamefont {Chen}},\
  }\href {\doibase 10.1103/PhysRevX.6.041006} {\bibfield  {journal} {\bibinfo
  {journal} {Phys. Rev. X}\ }\textbf {\bibinfo {volume} {6}},\ \bibinfo {pages}
  {041006} (\bibinfo {year} {2016})}\BibitemShut {NoStop}%
\bibitem [{\citenamefont {Levin}\ and\ \citenamefont
  {Stern}(2012)}]{levin2012kmatrix}%
  \BibitemOpen
  \bibfield  {author} {\bibinfo {author} {\bibfnamefont {M.}~\bibnamefont
  {Levin}}\ and\ \bibinfo {author} {\bibfnamefont {A.}~\bibnamefont {Stern}},\
  }\href {\doibase 10.1103/PhysRevB.86.115131} {\bibfield  {journal} {\bibinfo
  {journal} {Phys. Rev. B}\ }\textbf {\bibinfo {volume} {86}},\ \bibinfo
  {pages} {115131} (\bibinfo {year} {2012})}\BibitemShut {NoStop}%
\end{thebibliography}%

\end{document}